\pdfoutput=1 % only if pdf/png/jpg images are used

\documentclass[11pt,a4paper]{article}
\usepackage{maus-jinstpub}
\usepackage{url}
\usepackage{subcaption}
\usepackage{placeins}

\title{MAUS: The MICE Analysis User Software}

\author[a]{R. Asfandiyarov,}
\author[b]{R. Bayes,}
\author[c]{V. Blackmore,}
\author[d]{M. Bogomilov,} 
\author[c]{D. Colling,}  
\author[c]{A.J. Dobbs,} 
\author[a]{F. Drielsma,} 
\author[h]{M. Drews,} 
\author[c]{M. Ellis,}  
\author[e]{M. Fedorov,} 
\author[f]{P. Franchini,} 
\author[g]{R. Gardener,}
\author[f]{J.R. Greis,} 
\author[h]{P.M. Hanlet,} 
\author[i]{C. Heidt,}  
\author[c]{C. Hunt,}  
\author[h]{G. Kafka,} 
\author[a]{Y. Karadzhov,}  
\author[c]{A. Kurup,} 
\author[g]{P. Kyberd,}  
\author[g]{M. Littlefield,}  
\author[j]{A. Liu,}  
\author[c,n]{K. Long,}  
\author[k]{D. Maletic,}  
\author[c]{J. Martyniak,}  
\author[c]{S. Middleton,}  
\author[h,j]{T. Mohayai,}  
\author[g]{J.J. Nebrensky,}  
\author[b]{J.C. Nugent,}  
\author[l]{E. Overton,}  
\author[l]{V. Pec,}  
\author[f]{C.E. Pidcott,}  
\author[h,1]{D. Rajaram,\note{Corresponding author.}}  
\author[m]{M. Rayner,}  
\author[g]{I.D. Reid,}  
\author[n]{C.T. Rogers,}  
\author[c]{E. Santos,}
\author[k]{M. Savic,}  
\author[f]{I. Taylor,}  
\author[h]{Y. Torun,}  
\author[m]{C.D. Tunnell,}  
\author[c]{M.A. Uchida,}  
\author[a]{V. Verguilov,}  
\author[b]{K. Walaron,}  
\author[h]{M. Winter,}  
\author[l]{S. Wilbur}  
\affiliation[a]{DPNC, section de Physique, Universit\'e de Gen\`eve, Geneva, Switzerland}
\affiliation[b]{School of Physics and Astronomy, Kelvin Building, The University of Glasgow, Glasgow, UK}
\affiliation[c]{Department of Physics, Blackett Laboratory, Imperial College London, London, UK}
\affiliation[d]{Department of Atomic Physics, St. Kliment Ohridski University of Sofia, Sofia, Bulgaria}
\affiliation[e]{Radboud University of Nijmegen, Netherlands}
\affiliation[f]{Department of Physics, University of Warwick, Coventry, UK}
\affiliation[g]{Brunel University, Uxbridge, UK}
\affiliation[h]{Physics Department, Illinois Institute of Technology, Chicago, IL, USA}
\affiliation[i]{University of California, Riverside, CA, USA}
\affiliation[j]{Fermilab, Batavia, IL, USA}
\affiliation[k]{Institute of Physics, University of Belgrade, Serbia}
\affiliation[l]{Department of Physics and Astronomy, University of Sheffield, Sheffield, UK}
\affiliation[m]{Department of Physics, University of Oxford, Denys Wilkinson Building, Oxford, UK}
\affiliation[n]{STFC Rutherford Appleton Laboratory, Harwell Oxford, Didcot, UK}

\emailAdd{durga@fnal.gov}

\abstract{The Muon Ionization Cooling Experiment (MICE) collaboration has developed the MICE Analysis User Software (MAUS) to simulate and analyze experimental data.  It serves as the primary codebase for the experiment, providing for offline batch simulation and reconstruction as well as online data quality checks. The software provides both traditional particle-physics functionalities such as track reconstruction and particle identification, and  accelerator physics functions, such as calculating transfer matrices and emittances. The code design is object orientated, but has a top-level structure based on the Map-Reduce model. This allows for parallelization to support live data reconstruction during data-taking operations. MAUS allows users to develop in either Python or C++ and provides APIs for both. Various software engineering practices from industry are also used to ensure correct and maintainable code, including style, unit and integration tests, continuous integration and load testing, code reviews, and distributed version control. The software framework and the simulation and reconstruction capabilities are described.}


\begin{document}

\maketitle

\section{Introduction}\label{sec:intro}

%\linenumbers

\subsection{The MICE experiment} \label{sec:mice}
\begin{figure}[!htb]
\centering
\includegraphics[width=0.9\textwidth]{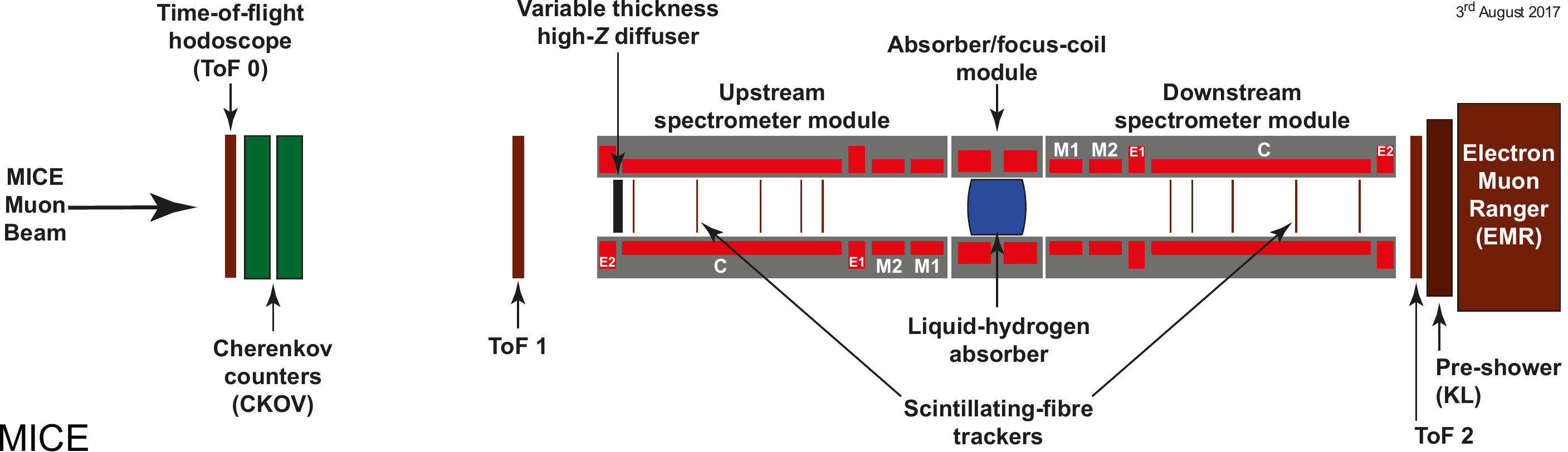}
\caption{Schematic diagram of the MICE experiment.The red rectangles represent the coils of the spectrometer solenoids and focus-coil module. The individual coils of the spectrometer solenoids are labelled E1, C, E2, M1 and M2. The various detectors (time-of-flight hodoscopes (TOF0, TOF1)~\cite{NIMA_TOF}, Cherenkov counters~\cite{CkovIEEE}, scintillating-fiber trackers~\cite{TrackersNIM}, KLOE-Light (KL) calorimeter~\cite{KLOE, KLOE2}, and Electron Muon Ranger (EMR)~\cite{EMRJINST, EMRJINST11}) are also represented.}
\label{fig:step4}
\end{figure}

The Muon Ionization Cooling Experiment (MICE) sited at the STFC Rutherford Appleton Laboratory (RAL) has delivered the first demonstration of muon ionization cooling~\cite{CoolingIPAC18} -- the reduction of the phase-space of muon beams. Muon-beam cooling is essential for future facilities based on muon acceleration, such as the Neutrino Factory or Muon Collider \cite{IDR, MC_Overview}. The experiment was designed to be built and operated in a staged manner. In the first stage, the muon beamline was commissioned ~\cite{BeamlineJINST} and characterized~\cite{BeamCharacterisationEurPhysJ}. A schematic diagram of the configuration used to study the factors that determine the performance of an ionization-cooling channel is shown in figure~\ref{fig:step4}.  The MICE experiment was operated such that muons passed through the experiment one at a time.  The experiment included instrumentation to identify particle species (the particle-identification detectors, PID)~\cite{NIMA_TOF, KLOE, KLOE2, PionContaminationJINST, EMRJINST, EMRJINST11} and to measure the phase-space coordinates of each muon.  An ensemble of muons that was representative of the muon beam was then assembled using the measured coordinates.  The techniques used to reconstruct the ensemble properties of the beam are described in~\cite{FirstSingleParticle} and the first observation of the ionization-cooling of a muon beam is presented in~\cite{CoolingIPAC18}.

The configuration shown in figure~\ref{fig:step4}  was used to study the factors that determine the performance of an ionization-cooling channel and to observe for the first time the reduction in transverse emittance of a muon beam.

The MICE Muon Beam line is described in detail in~\cite{BeamlineJINST}. There are 5 different detector systems present on the beamline: time-of-flight (TOF) scintillators \cite{NIMA_TOF}, threshold Cherenkov (Ckov) counters \cite{CkovIEEE}, scintillating-fiber trackers \cite{TrackersNIM}, a sampling calorimeter (KL) \cite{KLOE2, PionContaminationJINST}, and the Electron Muon Ranger (EMR) -- a totally active scintillating calorimeter~\cite{EMRJINST, EMRJINST11}. The TOF, Ckov, KL and EMR detectors are used for particle identification (PID), and the scintillating-fiber trackers are used to measure position and momentum. The TOF detector system consists of three detector stations, TOF0, TOF1 and TOF2, each composed of two orthogonal layers of scintillator bars.  The TOF system determines PID via the time-of-flight between the stations. Each station also provides a low-resolution image of the beam profile.  The Ckov system consists of two aerogel threshold Cherenkov stations, CkovA and CkovB. The KL and EMR detectors, the former using scintillating fibers embedded in lead sheets, and the latter scintillating bars, form the downstream calorimeter system.

The tracker system consists of two scintillating-fiber detectors, one upstream of the MICE cooling cell, the other downstream, in order to measure the change in emittance across the cooling cell. Each detector consists of 5 stations, each station having 3 fiber planes, allowing precision measurement of momentum and position to be made on a particle-by-particle basis.

\subsection{Software requirements} \label{sec:requirements}

The MICE software must serve both the accelerator-physics and the particle-physics needs of the experiment. Traditional particle-physics functionality includes reconstructing particle tracks, identifying them, and simulating the response from various detectors, while the accelerator-physics aspect includes the calculation of transfer matrices and Twiss parameters and propagating the beam envelopes. All of these items require a detailed description of the beamline, the geometries of the detectors, and the magnetic fields, as well as functionality to simulate the various detectors and reconstruct the detector outputs. MICE aims to measure the change in emittance to 1\%, which imposes requirements on the performance of the track reconstruction, particle identification and measurements of scattering widths. In addition, the computational performance of the software was also important in order to ensure that the software can reconstruct data with sufficient speed to support live online monitoring of the experiment.

\section{MAUS}\label{sec:maus}

The MICE  Analysis User Software (MAUS) is the collaboration's simulation, reconstruction, and  analysis software framework. MAUS provides a Monte Carlo (MC) simulation of the experiment, reconstruction of tracks and identification of particles from simulations and real data, and provides monitoring and diagnostics while running the experiment.

Installation is performed via a set of shell scripts with SCons~\cite{SCONS} as the tool for constructing and building the software libraries and executables. The codebase is maintained with GNU Bazaar~\cite{bazaar}, a distributed version control system, and is hosted on Launchpad~\cite{launchpad}, a website that provides functionalities to host and maintain the software repository. MAUS has a number of dependencies on standard packages such as Python, ROOT~\cite{ROOT} and Geant4~\cite{GEANT4} which are built as ``third party'' external libraries during the installation process.  The officially supported platform is Scientific Linux 6~\cite{scilinux} though developers have successfully built on CentOS~\cite{centos}, Fedora~\cite{fedora}, and Ubuntu~\cite{ubuntu} distributions.

Each of the MICE detector systems, described in section~\ref{sec:mice}, is represented within MAUS. Their data structures are described in section~\ref{sec:maus-datastr} and their simulation and reconstruction algorithms in sections~\ref{sec:mc} and \ref{sec:recon}. MAUS also provides ``global'' reconstruction routines, which combine data from individual detector systems to identify particle species by the likelihood method and perform a global track fit. These algorithms are also described in section~\ref{sec:recon}.

\subsection{Code design}\label{sec:maus-arch}

MAUS is written in a mixture of Python and C++. C++ is used for complex or low-level algorithms where processing time is important, while Python is used for simple or high-level algorithms where development time is a more stringent requirement. Developers are allowed to write in either Python or C++ and Python bindings to C++ are handled through internal abstractions. In practice, all the reconstruction modules are written in C++ but support is provided for legacy modules written in Python.

MAUS has an Application Programming Interface (API) that provides a framework on which developers can hang individual routines. The MAUS API provides MAUS developers with a well-defined environment for developing reconstruction code, while allowing independent development of the back-end and code-sharing of common elements, such as error handling. 

\begin{figure}[!htb]
\centering
\includegraphics[width=0.7\textwidth]{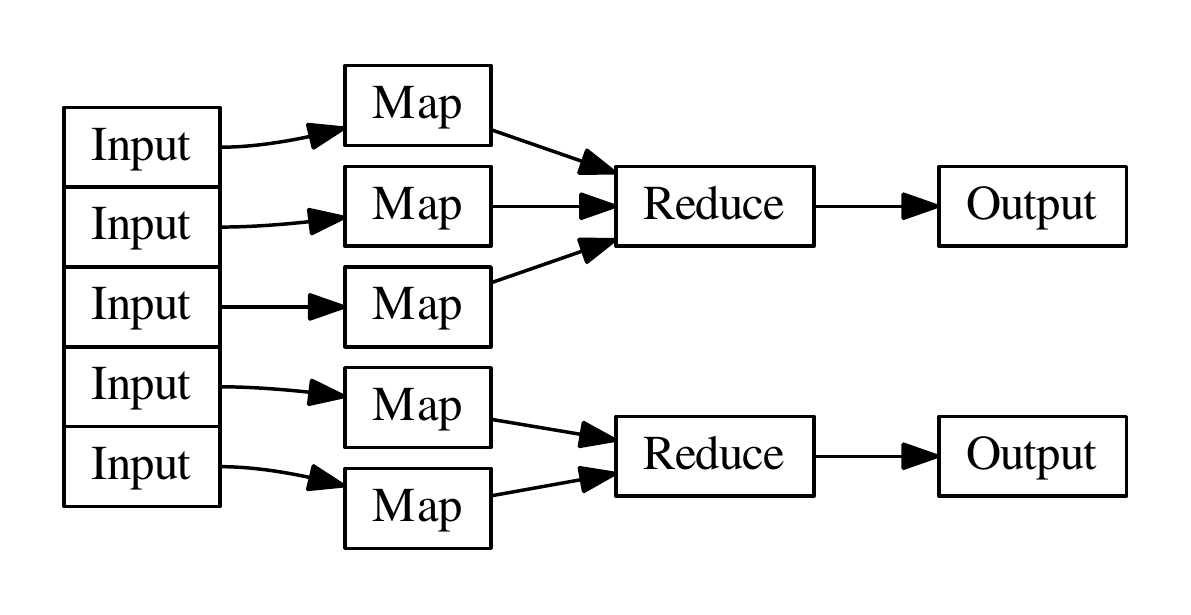}
\caption{A Map-Reduce framework.}
\label{fig:mapreduce}
\end{figure}

The MAUS data processing model is inspired by the Map-Reduce framework~\cite{MapReduce}, which forms the core of the API design. Map-Reduce, illustrated in figure~\ref{fig:mapreduce} is a useful model for parallelizing data processing on a large scale. A \textit{map} process takes a single object as an input, transforms it, and returns a new object as the output (in the case of MAUS this input object is the \emph{spill} class, see Section~\ref{sec:maus-datastr}).

A module is the basic building block of the MAUS API framework. Four types of module exist within MAUS:

\begin{enumerate}
\item \textbf{\textit{Inputters}} generate input data either by reading data from files or over a network, or by generating an input beam;
\item \textbf{\textit{Mappers}} modify the input data, for example by reconstructing signals from detectors, or tracking  particles to generate MC hits;
\item \textbf{\textit{Reducers}} collate the mapped data and provide functionality that requires access to the entire data set; and
\item \textbf{\textit{Outputters}} save the data  either by streaming over a network or writing to disk.
\end{enumerate}

\begin{figure}[!htb]
\centering
\includegraphics[width=1.0\textwidth]{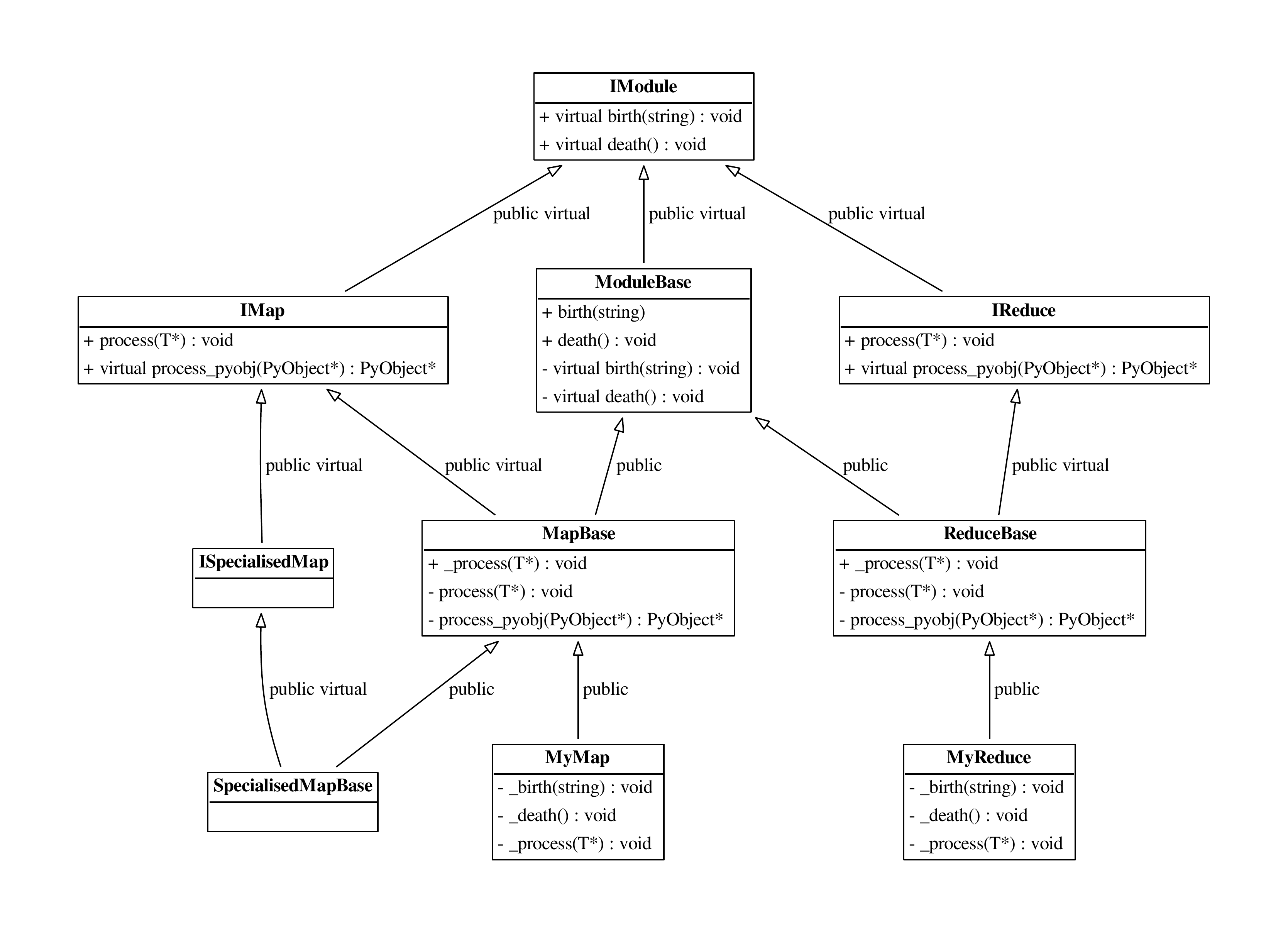}
\caption{The MAUS API class hierarchy for Map and Reduce modules. The input and output modules follow related designs. \emph{T} represents a templated argument. ``+'' indicates the introduction of a virtual void method, defining an interface, while ``-'' indicates that a class implements that method, fulfilling that aspect of the interface. The  \emph{process\_pyobj} functions are the main entry points for Python applications, and \emph{process} the entry points for C++ applications. The framework can be extended as many times as  necessary, as exemplified by the ``SpecialisedMap'' classes.}
\label{fig:api}
\end{figure}

\noindent Each module type follows a common, extensible, object-orientated class hierarchy, shown for the case of the \textit{map} and \textit{reduce} modules in figure~\ref{fig:api}. 

There are some objects that sit outside the scope of this modular framework but are nevertheless required by several of the modules. For instance,  the detector geometries, magnetic fields, and calibrations are required by the reconstruction and simulation modules, and objects such as the electronics-cabling maps are required in order to unpack data from the data acquisition (DAQ) source, and error handling functionality is required by all of the modules. All these objects are accessed through a static singleton \emph{globals} class. 

MAUS has two execution concepts. A \emph{job} refers to a single execution of the code, while a \emph{run} refers to the processing of data for a DAQ run or MC run. A job may contain many runs. Since data are typically accessed from a single source and written to a single destination, \textit{inputters} and \textit{outputters} are initialized and destroyed at the beginning and end of a job. On the other hand, \textit{mappers} and \textit{reducers} are initialized at the beginning of a run in order to allow run-specific information such as electronics cabling maps, fields, calibrations and geometries to be loaded.

The principal data type in MAUS, which is passed from module to module, is the \emph{spill}. A single spill corresponds to  data from the particle burst associated with a dip of the MICE target \cite{BeamlineJINST}. A spill lasts up to $\sim$ 3\,ms and contains several DAQ triggers.  Data from a given trigger define a single MICE \emph{event}. In the language of the Input-Map-Reduce-Output framework, an \textit{Input} module creates an instance of spill data, a \textit{Map} module processes the spill (simulating, reconstructing, etc.), a \textit{Reduce} module acts on a collection of spills when all the \textit{mappers} finish, and finally an \textit{Output} module records the data to a given file format.

Modules can exchange spill data either as C++ pointers or JSON~\cite{JSON} objects. In Python, the data format can be changed by using a converter module, and in C++ \textit{mappers} are templated to a MAUS data type and an API  handles any necessary conversion to that type (see Fig.~\ref{fig:api}).

Data contained within the MAUS data structure (see Section~\ref{sec:maus-datastr}) can be saved to permanent storage in one of two formats. The default data format is a ROOT~\cite{ROOT} binary and the secondary format is JSON. ROOT is a standard high-energy physics analysis package, distributed with MAUS, through which many of the analyses on MICE are performed. Each spill is stored as a single entry in a ROOT TTree object.  JSON is an ASCII data-tree format. Specific JSON parsers are available -- for example, the Python \emph{json} library, and the C++ \emph{JsonCpp} \cite{JSONCPP} parser come prepackaged with MAUS. 

In addition to storing the output from the \textit{map} modules, MAUS is also capable of storing the data  produced by  \textit{reducer} modules using a special \emph{Image} class. This class is used by \emph{reducers} to store images of monitoring histograms, efficiency plots, etc. \emph{Image} data may only be saved in JSON format.

\subsection{Data structure}\label{sec:maus-datastr}

\subsubsection{Physics data} \label{sec:physics-datastr}

\begin{figure}[!p]
\centering
\includegraphics[width=\textwidth]{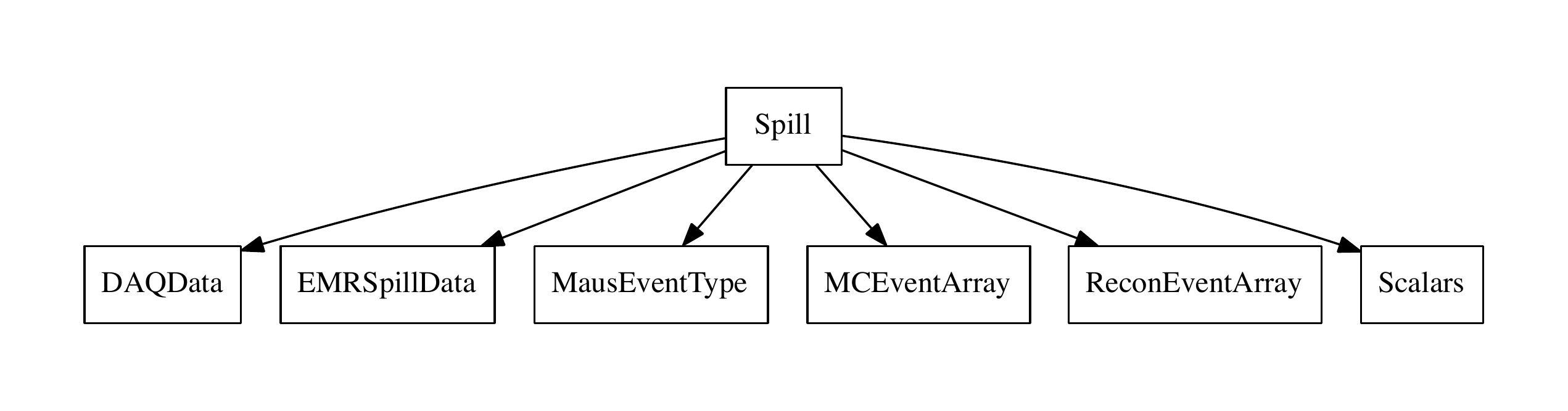}
\caption{The MAUS output structure for a spill event. The label in each box is the name of the C++ class.}
\label{fig:datastructure-spill}
\end{figure}

\begin{figure}[ptb]
\centering
\includegraphics[width=\textwidth]{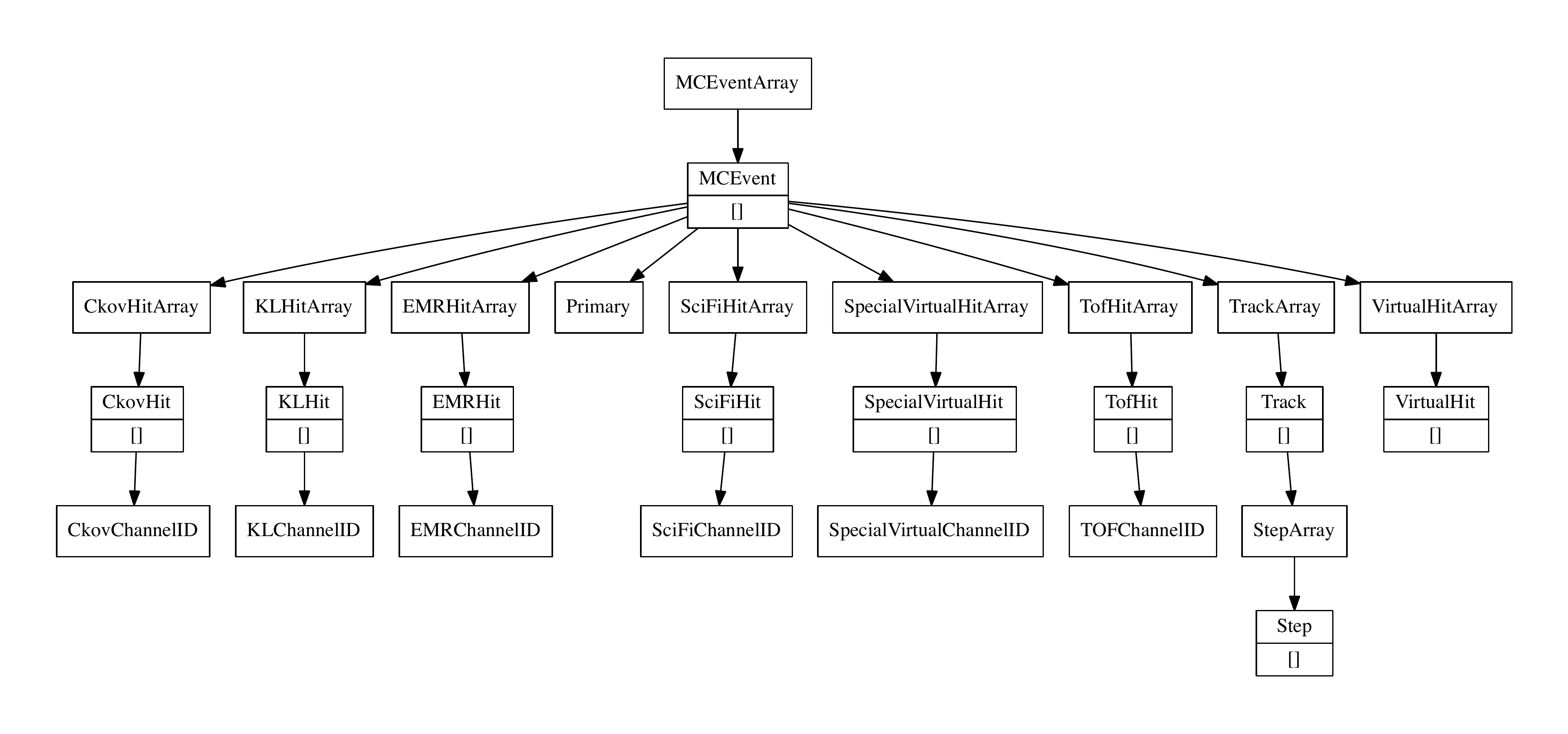}
\caption{The MAUS data structure for MC events. The label in each box is the name of the C++ class and [] indicates that child objects are array items.}
\label{fig:datastructure-mc}
\end{figure}

\begin{figure}[ptb]
\centering
\includegraphics[width=\textwidth]{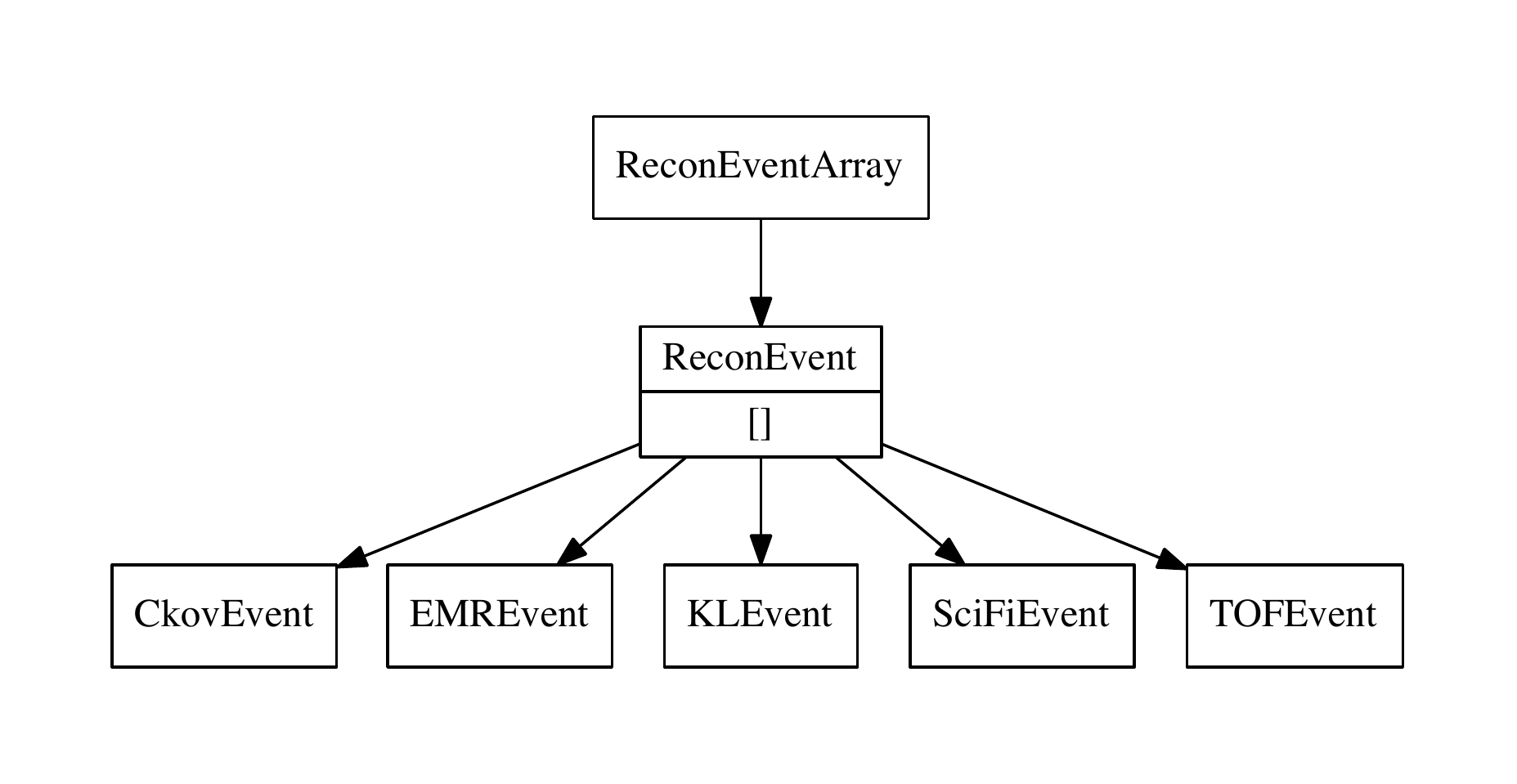}
\caption{The MAUS data structure for reconstructed events. The label in each box is the name of the C++ class.}
\label{fig:datastructure-recon}
\end{figure}

\begin{figure}[ptb]
\centering
\includegraphics[width=0.3\textwidth]{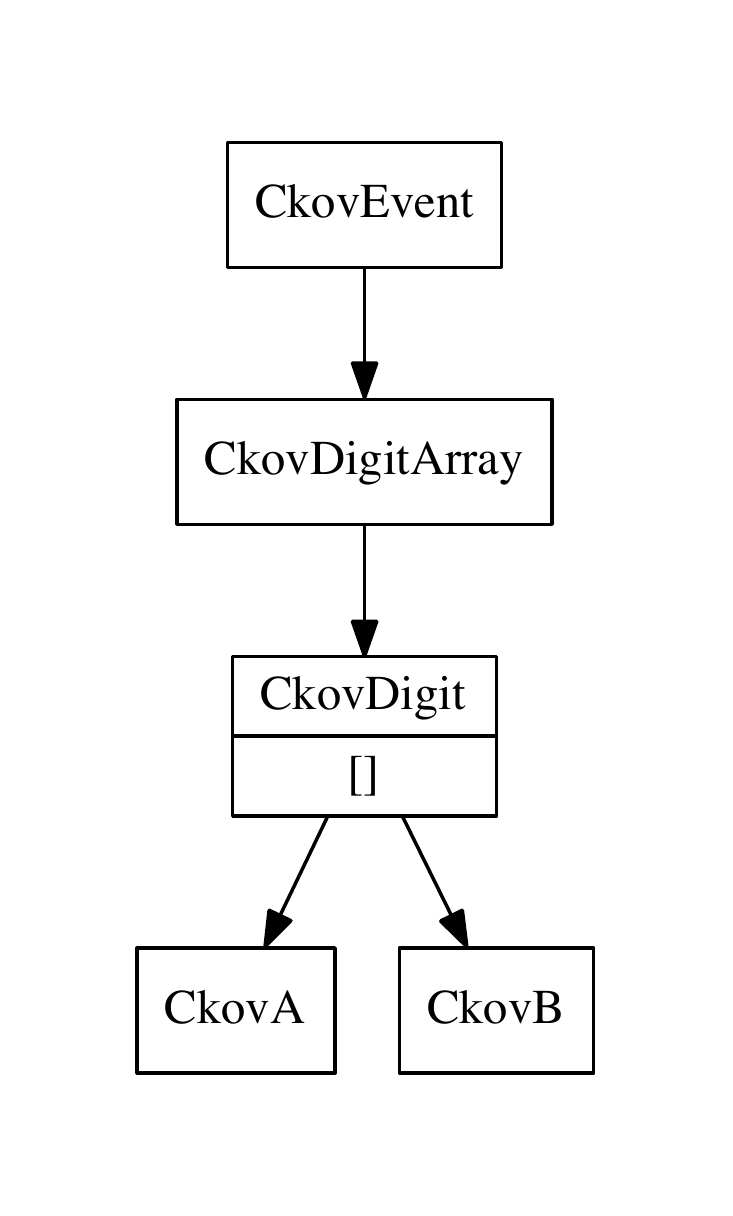}
\includegraphics[width=0.25\textwidth]{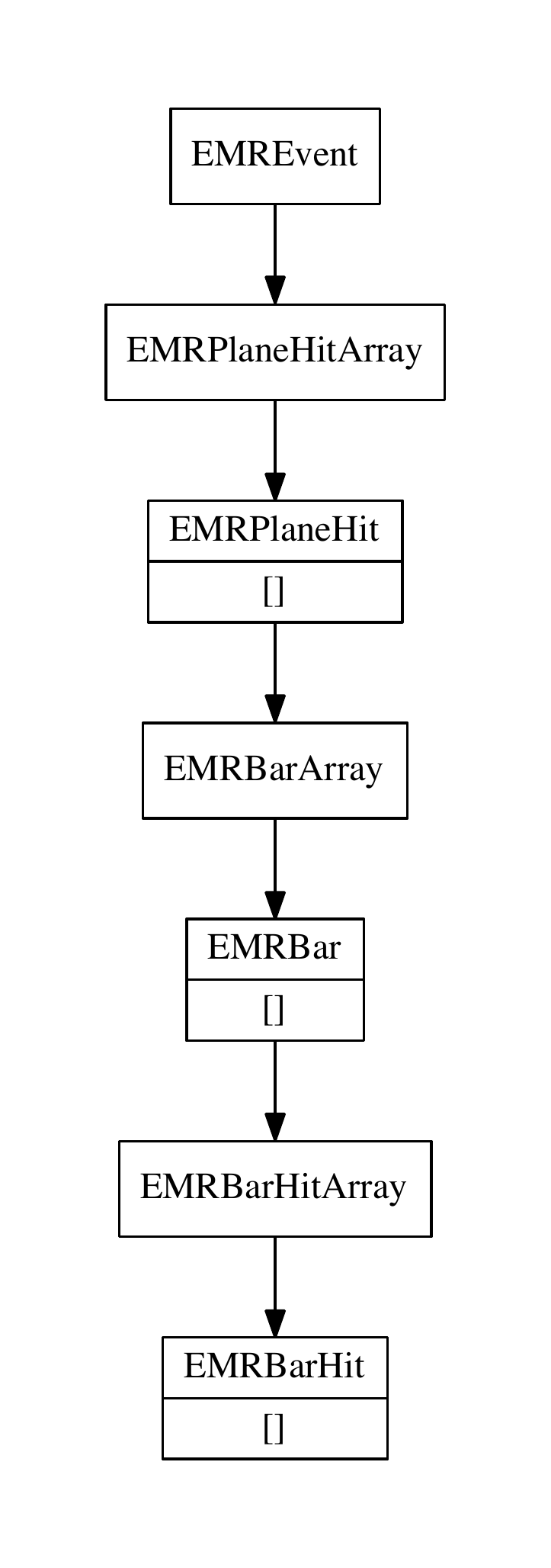}
\includegraphics[width=0.4\textwidth]{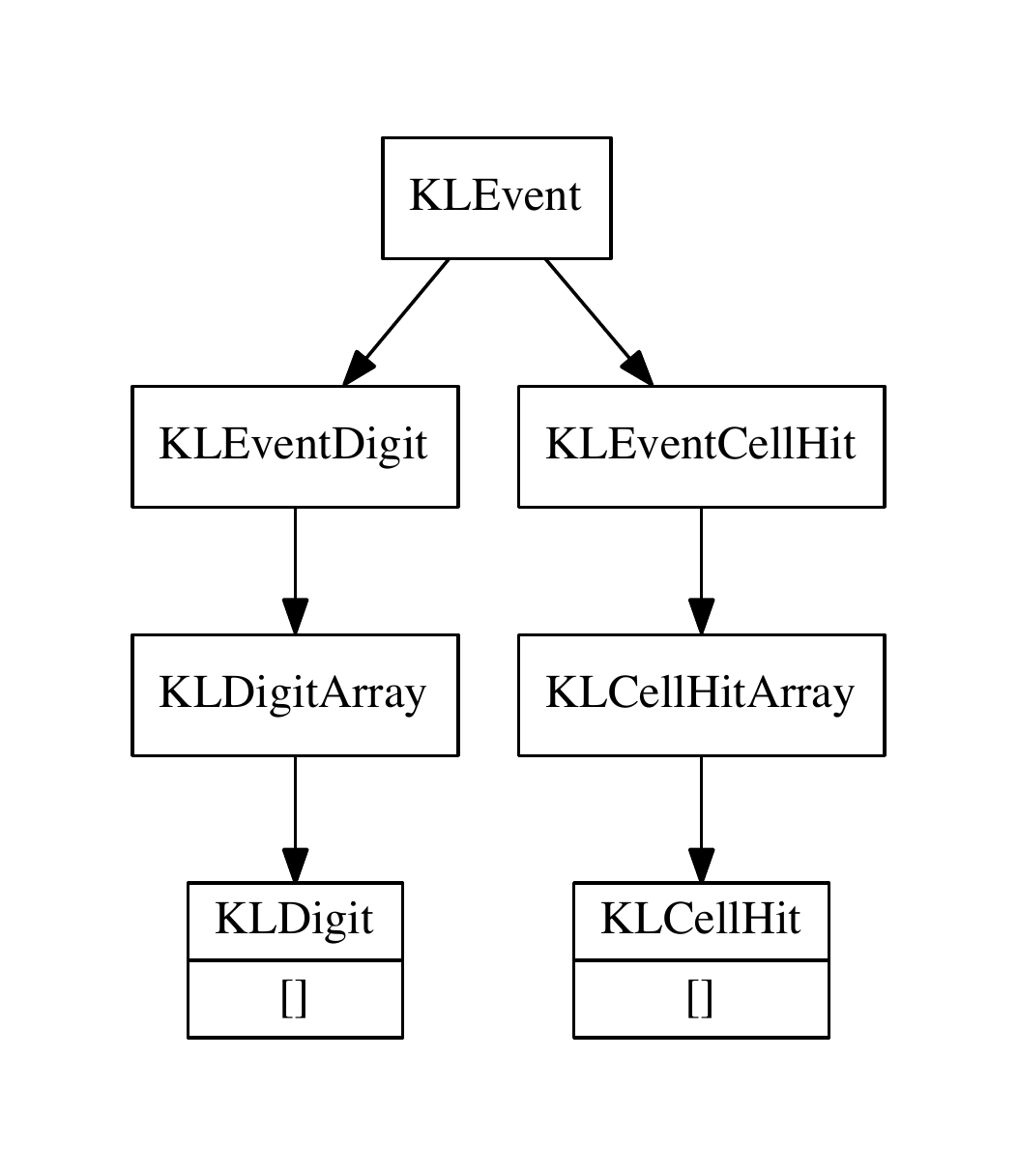}
\caption{The MAUS data structure for CKOV (left), EMR (middle) and KL (right) reconstructed events. The label in each box is the name of the C++ class [] indicates that child objects are array items.}
\label{fig:datastructure-recon-ckov-emr-kl}
\end{figure}

\begin{figure}[ptb]
\centering
\includegraphics[width=1.0\textheight,angle=90,origin=c]{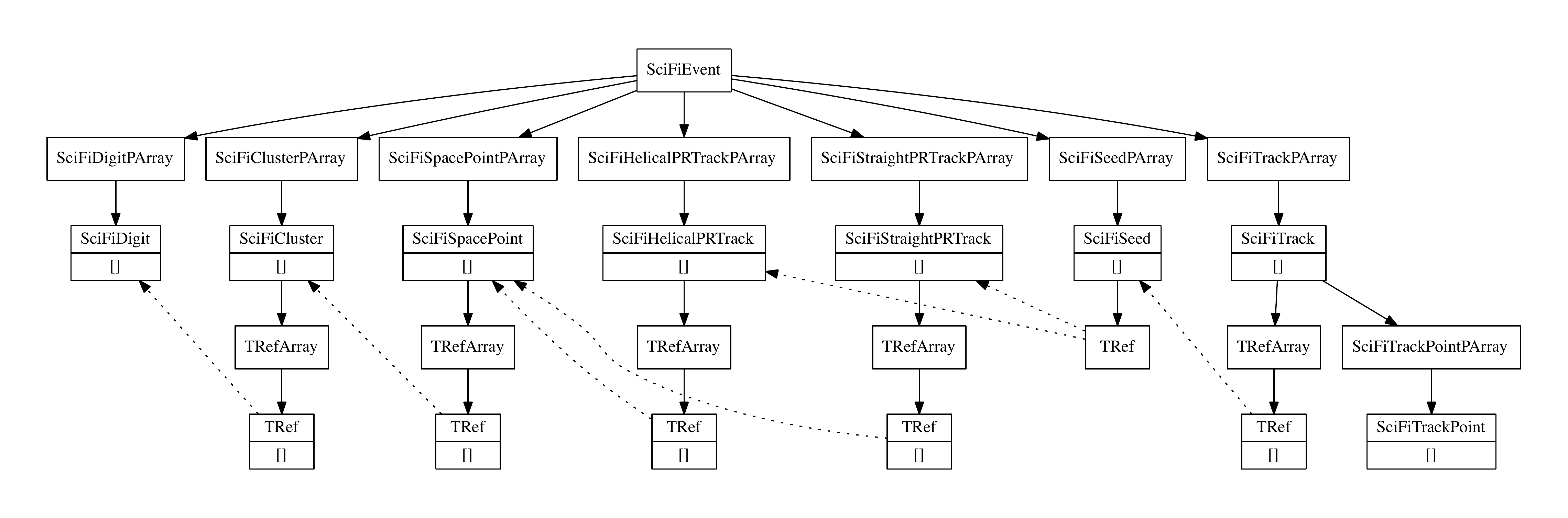}
\caption{The MAUS data structure for the tracker. The label in each box is the name of the C++ class and []  indicates that child objects are array items.}
\label{fig:datastructure-recon-scifi}
\end{figure}

\begin{figure}[ptb]
\centering
\includegraphics[width=1.0\textheight,angle=90,origin=c]{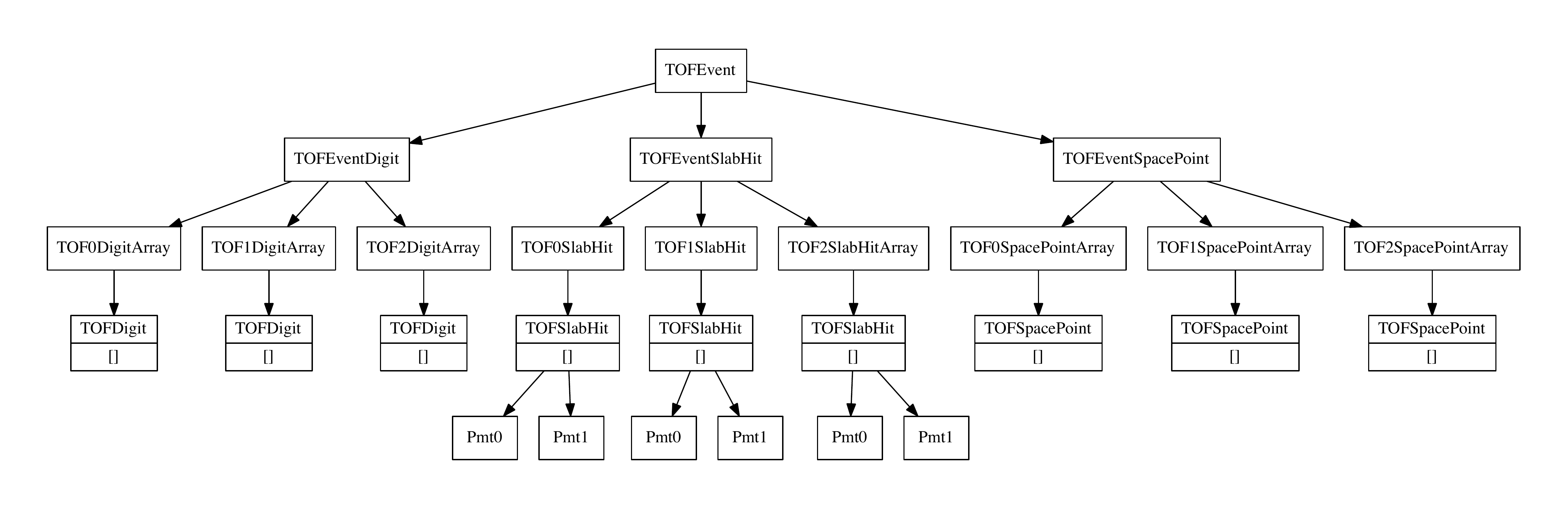}
\caption{The MAUS data structure for the TOFs. The label in each box is the name of the C++ class and [] indicates that child objects are array items.}
\label{fig:datastructure-recon-tof}
\end{figure}

At the top of the MAUS data structure is the spill class which contains all the data from the simulation, raw real data and the reconstructed data. The spill is passed between modules and written to permanent storage. The data within a spill is organized into arrays of three possible event types: an \emph{MCEvent} contains data  representing the simulation of a single particle traversing the experiment and the simulated detector responses; a \emph{DAQEvent} corresponds to the real data for a single trigger; and a \emph{ReconEvent} corresponds to the data reconstructed for a single particle event (arising either from a Monte Carlo(MC) particle or a real data trigger). These different branches of the MAUS data structure are shown diagrammatically in figures~\ref{fig:datastructure-spill}--\ref{fig:datastructure-recon-tof}.

The sub-structure of the MC event class is shown in figure~\ref{fig:datastructure-mc}. The class is subdivided into events containing detector hits (energy deposited, position, momentum) for each of the MICE detectors (see Section~\ref{sec:mice}). The event also contains information about the primary particle that created the hits in the detectors.

The sub-structure of the reconstruction event class is shown in figure~\ref{fig:datastructure-recon}. The class is subdivided into events representing each of the MICE detectors, together with the data from the trigger, and data for the global event reconstruction. Each detector class and the global-reconstruction class has several further layers of reconstruction data. This is shown in figures~\ref{fig:datastructure-recon-ckov-emr-kl}--\ref{fig:datastructure-recon-tof}.

\FloatBarrier

\subsubsection{Top level data organization} \label{sec:top-level-datastr}

In addition to the spill data, MAUS also contains structures for storing supplementary information for each run and job. These are referred to as \emph{JobHeader} and \emph{JobFooter}, and \emph{RunHeader} and \emph{RunFooter}. The \emph{JobHeader} and \emph{JobFooter} represent data, such as the MAUS release version, associated with the start and end of a job, and the \emph{RunHeader} and \emph{RunFooter} represent data, such as the geometry and calibrations associated with a run, associated with the start and end of a run. These are saved to the output along with the spill.

\begin{figure}[htb]
\centering
\includegraphics[width=1.03\textwidth]{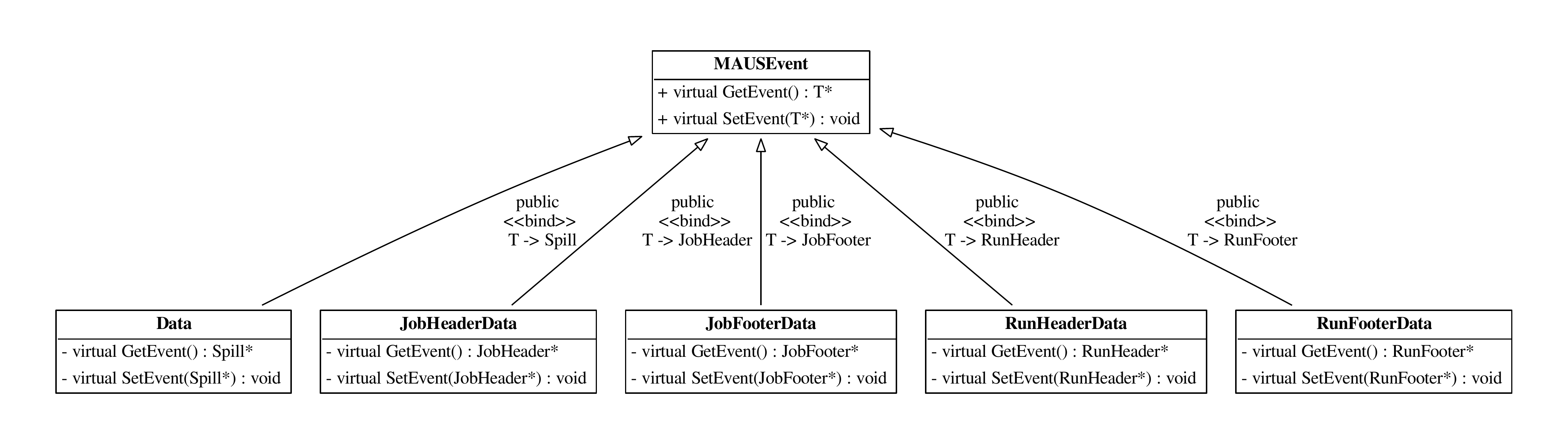}
\caption{Class hierarchy for the wrappers and base class of the top-level classes of the MAUS data structure.}
\label{fig:top-level}
\end{figure}

In order to interface with ROOT, particularly in order to save data in the ROOT format, thin wrappers for each of the top level classes, and a templated base class, were introduced. This allows the ROOT TTree, in which the output data is stored (see Section~\ref{sec:physics-datastr}), to be given a single memory address to read from. The wrapper for Spill is called \emph{Data}, while for each of RunHeader, RunFooter, JobHeader and JobFooter, the respective wrapper class is just given the original class name with ``Data'' appended, e.g., \emph{RunHeaderData}. The base class for each of the wrappers is called \emph{MAUSEvent}. The class hierarchy is illustrated in Figure~\ref{fig:top-level}.

\subsection{Data flow}\label{sec:maus-dataflow}

\begin{figure}[htbp]
  \centering
  \includegraphics[width=0.9\textheight, angle=90, origin=c]{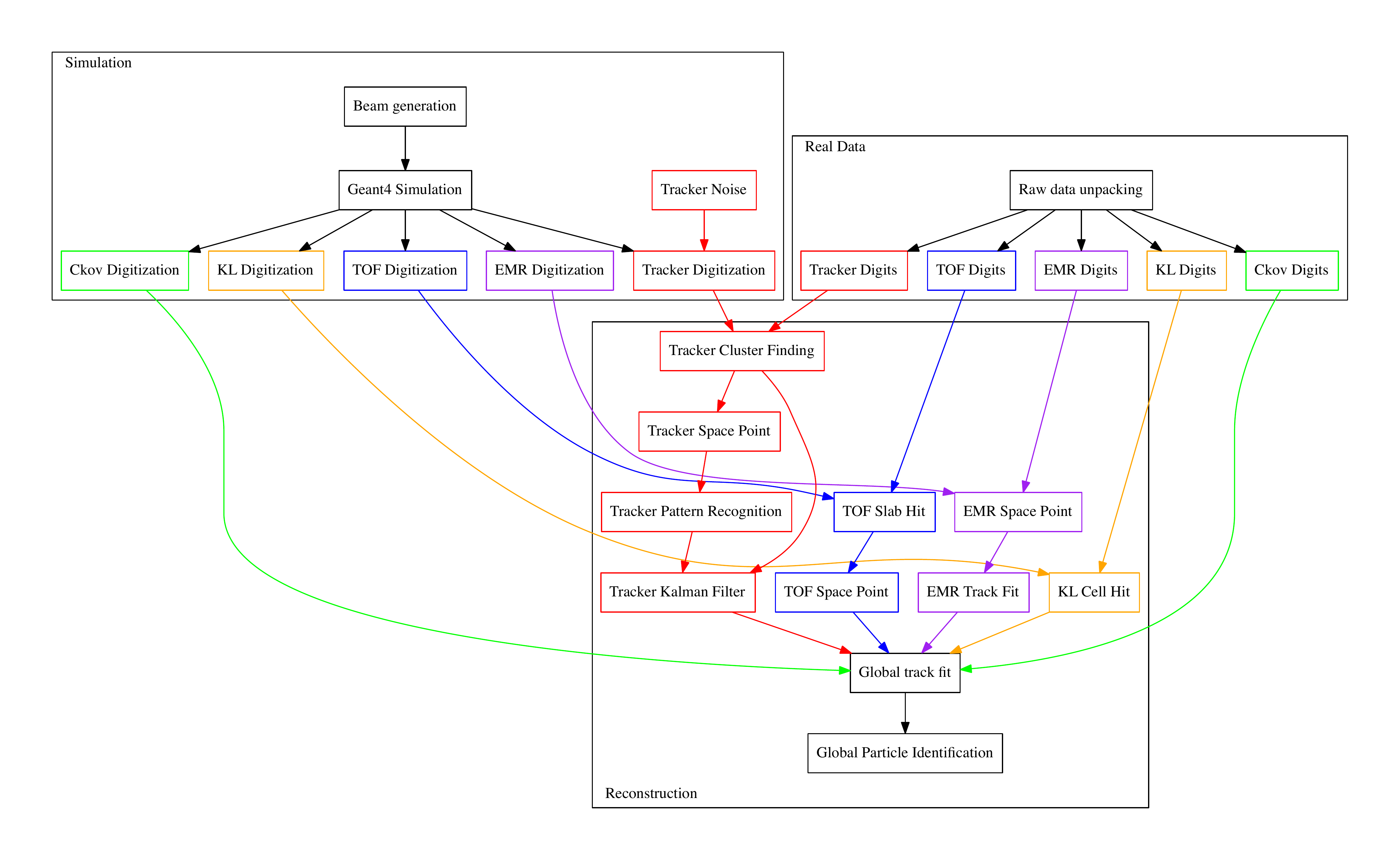}
  \caption{Data flow for the MAUS project. The data flow is color-coded by detector: Ckov - green, EMR - purple, KL - orange, TOF - blue, Tracker - red. }
  \label{fig:maus_process_diagram}
\end{figure}

The MAUS data-flow, showing the reconstruction chain for data originating from MC or real data, is depicted in figure~\ref{fig:maus_process_diagram}. Each item in the diagram is implemented as an individual module. The data flow is grouped into three principal areas: the simulation data flow used to generate digits (electronics signals) from particle tracking; the real data flow used to generate digits from real detector data; and the reconstruction data flow which illustrates how digits are built into higher level objects and converted to parameters of interest. The reconstruction data flow is the same for digits from real data and simulation.  In the case of real data, separate input modules are provided to read either directly from the DAQ, or from archived data stored on disk. A \textit{reducer} module for each detector provides functionality to create summary histograms.

\subsection{Testing}\label{sec:maus-tests}
MAUS has a set of tests at the unit level and the integration level, together with code-style tests for both Python and C++. Unit tests are implemented to test a single function, while integration tests operate on a complete workflow. Unit tests check that each function operates as intended by the developer. Tests are run automatically for every version committed to the repository and results show that a high level of code coverage has been achieved. Integration tests allow the overall performance of the code to be checked against specifications. The MAUS team provides unit test coverage that executes 70--80 $\%$ of the total code base. This level of coverage typically results in a code that performs the major workflows without any problems.

The MAUS codebase is built and tested using a Jenkins~\cite{Jenkins} continuous integration environment deployed on a cluster of servers. Builds and tests of the development branch are automatically triggered when there is a change to the codebase.  Developers are asked to perform a build and test on a personal branch of the codebase using the test server before requesting a merge with the development trunk. This enables the MAUS team to make frequent clean releases. Typically MAUS works on a 4--8 week major-release cycle.

\section{Monte Carlo}\label{sec:mc}
The Monte Carlo simulation of MICE encompasses beam generation, geometrical description of detectors  and fields, tracking of particles through detectors and digitization of the detectors' response to particle interactions. 

\subsection{Beam generation}\label{sec:beam}
Several options are provided to generate an incident beam.  Routines are provided to sample particles from a multivariate Gaussian distribution or generate ensembles of identical particles (pencil beams). In addition, it is possible to produce time distributions that are either rectangular or triangular in time to give a simplistic representation of the MICE time distribution. Parameters, controlled by data-cards, are available to control random seed generation,  relative weighting of particle species and the transverse-to-longitudinal coupling in the beam. MAUS also allows the generation of a polarized beam. 

Beam particles can also be read in from an external file created by G4Beamline~\cite{G4Beamline} -- a particle-tracking  simulation program based on Geant4, or ICOOL~\cite{ICOOL} -- a simulation program that was developed to study the ionization cooling of muon beams, as well as files in user-defined formats. In order to generate beams which are more realistic taking into account the geometry and fields of the actual MICE beamline, we use G4Beamline to model the MICE beamline from the target to a point upstream of the second quad triplet (upstream of Q4).  The beamline settings, \textit{e.g.,} magnetic field strengths and number of particles to generate, are controlled through data-cards. The magnetic field strengths have been tuned to produce beams that are reasonably accurate descriptions of the real beam. Scripts to install G4beamline are shipped with MAUS. 

Once the beam is generated, the tracking and interactions of particles as they traverse the rest of the beamline and the MICE detectors  are performed using Geant4.

\subsection{Geant4}\label{sec:geant}
\begin{figure}[htbp]
  \centering
  \includegraphics[width=0.9\textwidth]{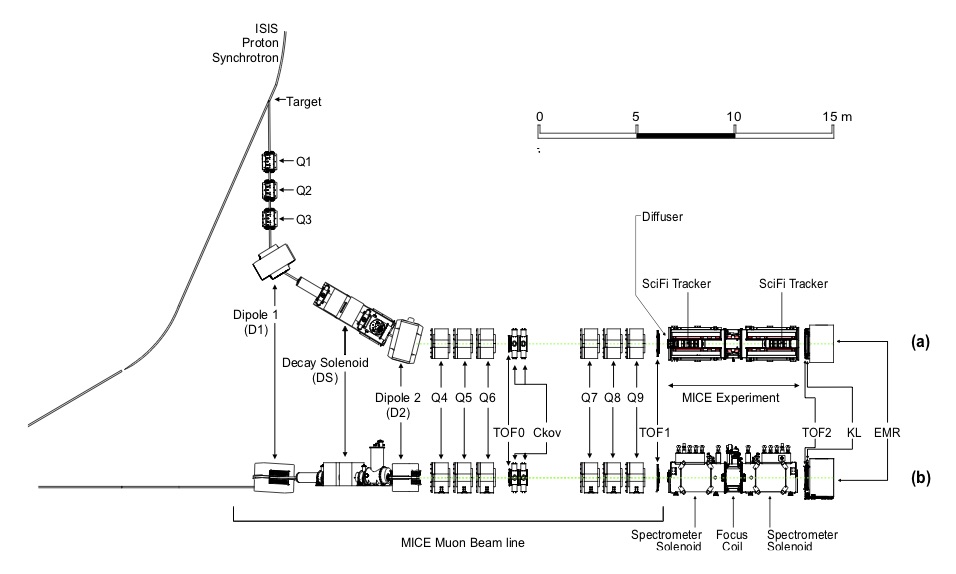}
  \caption{(a) Top and (b) side views of the MICE Muon Beamline, its instrumentation, and the experimental configuration. A titanium target dipped into the ISIS proton synchrotron and the resultant spill of particles was captured with a quadrupole triplet (Q1--3) and transported through momentum-selecting dipoles (D1, D2). The quadrupole triplets (Q4--6, Q7--9) transported particles to the upstream spectrometer module. The time-of-flight of particles, measured between TOF0 and TOF1, was used for particle identification.}
  \label{fig:beamline}
\end{figure}

A drawing of the MICE Muon Beam line~\cite{BeamlineJINST} is shown in figure~\ref{fig:beamline}. It consists of a quadrupole triplet (Q123) that captures pions produced when the MICE target intersects the ISIS proton beam, a pion-momentum-selection dipole (D1), a superconducting solenoid (DS) to focus and transport the particles to a second dipole (D2) that is used to select the muon-beam momentum, and a transport channel composed of a further two quadrupole triplets (Q456 and Q789). As described in the next section, the positions and apertures of the beamline magnets were surveyed and are reproduced in the geometry along with windows and materials in the path of the muon beams. The Geant4 simulation within MAUS starts 1\,m downstream of the second beamline dipole magnet D2. Geant4 bindings are encoded in the Simulation module. Geant4 groups particles by run, event and track. A Geant4 run maps to a MICE spill; a Geant4 event maps to a single inbound particle from the beamline; and a Geant4 track corresponds to a single particle in the experiment.\\

Geant4 provides a variety of reference physics processes to model the interactions of particles with matter. The default process in MAUS is ``\emph{QGSP\_BERT}'' which causes Geant4 to model hadron interactions using a Bertini cascade model up to 10 GeV/$c$~\cite{GEANT4HPhysics}. MAUS provides methods to set up the Geant4 physical processes through user-controlled data-cards. Finally, MAUS provides routines to extract particle data from the Geant4 tracks at user-defined locations.

\subsection{Geometry}\label{sec:geo}

MAUS uses an online Configuration Database to store all of its
geometries. These geometries have been extracted from CAD drawings
which are  updated based on the most recent surveys and technical drawings
available. The CAD drawings are translated to a geometry-specific
subset of XML, the Geometry Description Markup Language (GDML)~\cite{GDML} prior
to being recorded in the configuration database through the use of the 
FastRAD~\cite{fastrad} commercial software package. 

The GDML formatted description contains the beamline elements and the positions of
the detector survey points. Beam-line elements are described using 
tessellated solids to define the shapes of the physical
volumes. The detectors themselves are described using an independently
generated set of GDML files using Geant4 standard volumes. An
additional XML file is appended to the geometry description that
assigns magnetic fields and associates the detectors to their
locations in the GDML files. This file is
initially written by the geometry maintainers and formatted to contain
run-specific information during download.

The GDML files can be read via a
number of  libraries in Geant4 and ROOT for the
purpose of independent validation. The files are in turn translated into the MAUS-readable
geometry files either by accessing directly the data using a python
extension or through the use of EXtensible Stylesheet Language Transformations
(XSLT)~\cite{xslt}.

\subsection{Tracking, field maps and beam optics}\label{sec:fieldoptics}
MAUS tracking is performed using Geant4. By default, MAUS uses 4$^\mathrm{th}$ order Runge-Kutta (RK4) for tracking, although other routines are available. RK4 has been shown to have very good precision relative to the MICE detector resolutions, even for step sizes of several cm.

In a solenoid focussing lattice a cylindrically symmetric beam can be described by the 4D RMS beam emittance $\varepsilon_N$ and optical parameters $\beta_\perp$ and $\beta'_\perp$, its derivative with respect to z. $\beta_\perp$ is related to the variance of the position of particles $x$ by~\cite{PennEnvelope}:
\begin{equation}
\beta_\perp = \frac{p_{z}\mathrm{Var(}x\mathrm{)}}{\varepsilon_N mc};
\end{equation}
where $m$ is the particle mass, $c$ is the speed of light, and $p_z$ is the beam longitudinal momentum. In the approximation that particles travel near to the solenoid axis, transport of the beam envelope can be performed by integration of the differential equation:
\begin{equation}
2\beta_\perp \beta_\perp'' - (\beta_\perp')^2 + 4 \beta_\perp^2 \kappa^2 - 4 (1+\mathcal{L})^2 = 0.
\end{equation}
Transport of individual particles can be performed using numerical integration of the Lorentz force law. Alternately transport can be performed by calculating a transfer map $\mathbf{M}$ defined by:
\begin{equation}
\vec{u}_{ds} = \mathbf{M} \vec{u}_{us};
\end{equation}
where $\vec{u}_{us}$ and $\vec{u}_{ds}$ are the upstream and downstream transverse phase space vectors $\vec{u} = (x, p_x, y, p_y)$. MAUS can calculate the transfer map at arbitrary order by transporting a handful of particles and fitting to a multidimensional polynomial in $\vec{u}$.

Electromagnetic field maps are implemented in a series of overlapping regions. The world volume is divided into a number of voxels, and the field maps that impinge on each voxel is stored in a list. At each tracking step, MAUS iterates over the list of fields that impinge on the voxels within which the particle is stepping. For each field map, MAUS transforms to the local coordinate system of the field map, and calculates the field. The field values are transformed back into the global coordinate system, summed, and passed to Geant4. The voxelization enables the simulation of long accelerators without a performance penalty.

Numerous field types have been implemented within the MAUS framework. Solenoid fields can be calculated numerically from cylindrically symmetric 2D field maps, by taking derivatives of an on-axis solenoidal field or by using the sum of fields from a set of cylindrical current sheets. The use of field maps enables the realistic reproduction of the MICE apparatus, while a derivatives-based approach enables the exclusion of different terms in the higher order parts of the transfer map~\cite{NonlinearNote}. Multipole fields can be calculated from a 3D field map, or by taking derivatives from the usual multipole expansion formulae. Linear, quadratic and cubic interpolation routines have been implemented for field maps. Pillbox fields can be calculated by using the Bessel functions appropriate for a TM010 cavity or by reading a cylindrically symmetric field map.

The transport algorithms have been compared with each other and experimental data and show agreement at linear order~\cite{MiddletonThesis} in $\vec{u}$. Work is ongoing to study the effect of aberrations in the optics, indicated by non-linear terms in the transfer map relationship. These aberrations can cause distortion of the beam leading to emittance growth, which has been observed in the tails of the MICE beam. The tracking in MAUS has been benchmarked against ICOOL, G4Beamline, and MaryLie~\cite{MaryLie}, demonstrating good agreement. The routines have been used to model a number of beamlines and rings, including a neutrino factory front-end ~\cite{MAUSNuFact}.

\subsection{Detector response and digitization}\label{sec:detresp}
The modeling of the detector response and electronics enables MAUS to provide  data used to test reconstruction algorithms and estimate the uncertainties introduced by detectors and their readout.

The interaction of particles in materials is modeled using Geant4.  For each detector, a ``sensitive detector'' class processes Geant4 hits in active detector volumes and stores hit information such as the volume that was hit, the energy deposited and the time of the hit. Each detector's digitization routine then simulates the response of the electronics to these hits, modeling processes such as the photo-electron yield from a scintillator bar, attenuation in light guides and the pulse shape in the electronics. The data structure of the outputs from the digitizers are designed to match the output from the unpacking of real data from the DAQ.

\section{Reconstruction}\label{sec:recon}
The reconstruction chain takes as its input either digitized hits from the MC or DAQ digits from real data. Regardless, the detector reconstruction algorithms, by requirement and design, operate the same way on both MC and real data. 

\subsection{Time of flight}

There are three time-of-flight detectors in MICE which serve to distinguish particle type. The detectors are made of plastic scintillator and in each station there are orthogonal $x$ and $y$ planes with 7 or 10 slabs in each plane. 

Each Geant4 hit in the TOF is associated with a physical scintillator slab. The energy deposited by a hit is first converted to units of photo-electrons. The photo-electron yield from a hit accounts for the light attenuation corresponding to the distance of the hit from the photomultiplier tube (PMT) and is then smeared by the photo-electron resolution. The yields from all hits in a given slab are then summed and the resultant yield is converted to ADC counts.

The time of the hit in the slab is propagated to the PMTs at either end of the slab. The propagated time is then smeared by the PMT's time resolution and converted to TDC counts. Calibration corrections based on real data are then added to the TDC values so that, at the reconstruction stage, they can be corrected just as is done with real data.

The reconstruction proceeds in two main steps. First, the slab-hit-reconstruction takes individual PMT digits and associates them to reconstruct the hit in the slab. If there are multiple hits associated with a PMT, the hit which is earliest in time is taken to be the real hit. Then, if both PMTs on a slab have fired, the slab is considered to have a valid hit. The TDC values are converted to time and the hit time and charge associated with the slab hit are taken to be the average of the two PMT times and charges respectively. In addition, the product of the PMT charges is also calculated and stored. Secondly, individual slab hits are used to form space-points. A space-point  in the TOF is a combination of $x$ and $y$ slab hits. All combinations of $x$ and $y$ slab hits in a given station are treated as space-point candidates. Calibration corrections, stored in the Configurations Database, are applied to these hit times and if the reconstructed space-point is consistent with the resolution of the detector, the combination is said to be a valid space-point. The TOF has been shown to provide good time resolutions at the 60 ps level~\cite{NIMA_TOF}.

\begin{figure}[htbp]
  \centering
    \includegraphics[width=0.75\textwidth]{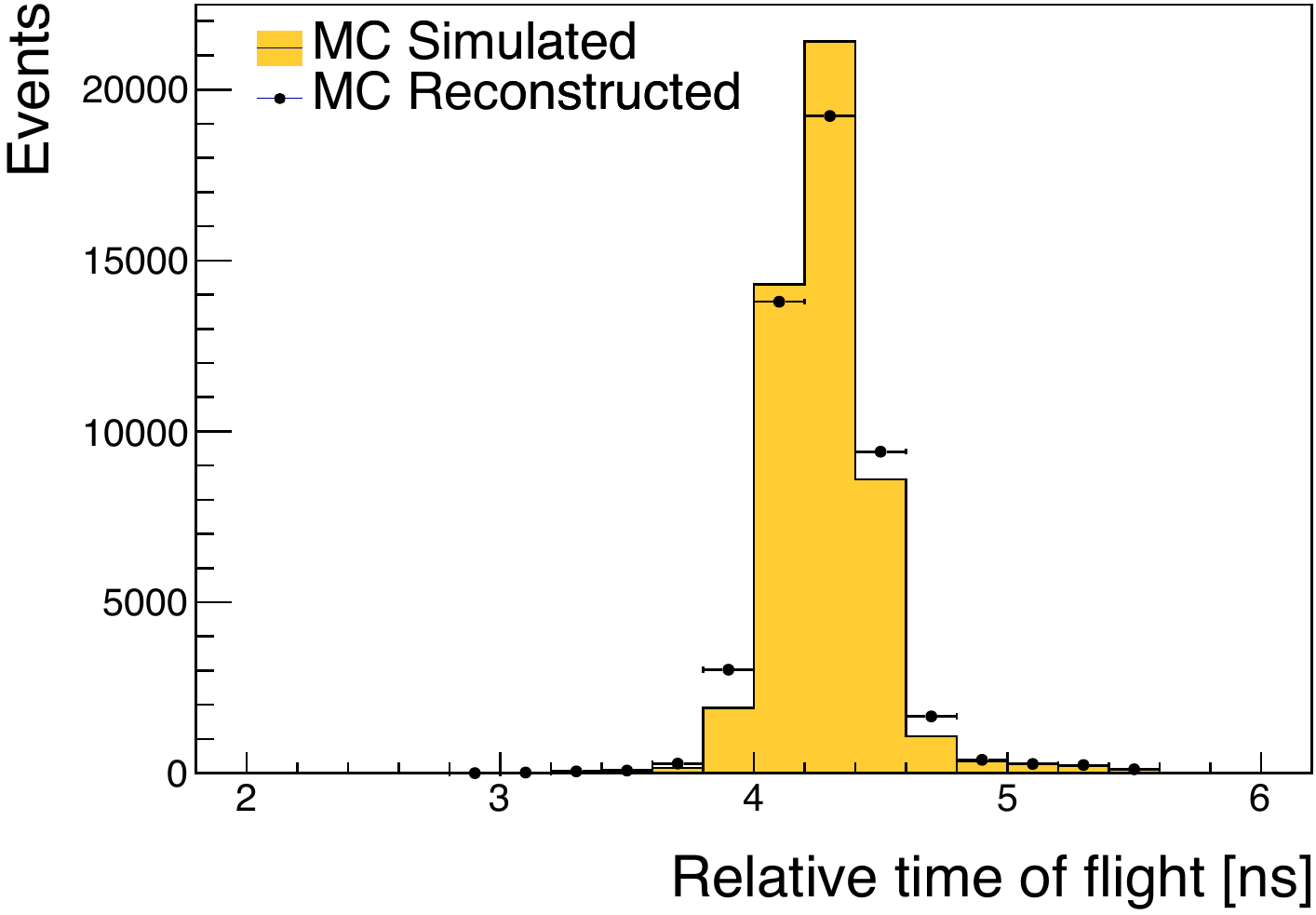}
  \caption{Relative time of flight between TOF0 and TOF1. The yellow histogram represents true MC events and the solid markers represent the same sample reconstructed with MAUS.}
  \label{fig:tof01t}
\end{figure}

\subsection{Scintillating-fiber trackers}

The scintillating-fiber trackers are the central piece of the reconstruction. As mentioned in Section~\ref{sec:mice}, there are two trackers, one upsteam and the other downstream of an absorber, situated within solenoidal magnetic fields. The trackers measure the emittance before and after particles pass through the absorber.

The tracker software algorithms and performance are described in detail in \cite{TrackerSoftwareJINST}. Digits are the most basic unit fed into the main reconstruction module, each digit representing a signal from one channel. Digits from adjacent channels are assumed to come from the same particle and are grouped to form clusters. Clusters from channels which intersect each other, in at least two planes from the same station, are used to form space-points, giving $x$ and $y$ positions where a particle intersected a station. Once space-points have been found, they are associated with individual tracks through pattern recognition (PR), giving straight or helical PR tracks. These tracks, and the space-points associated with them, are then sent to the final track fit. To avoid biases that may come from space-point reconstruction, the Kalman filter uses only reconstructed clusters as input.

\begin{figure*}[t!]
\centering
\begin{subfigure}[t]{0.495\textwidth}
\centering
\includegraphics[width=0.99\textwidth]{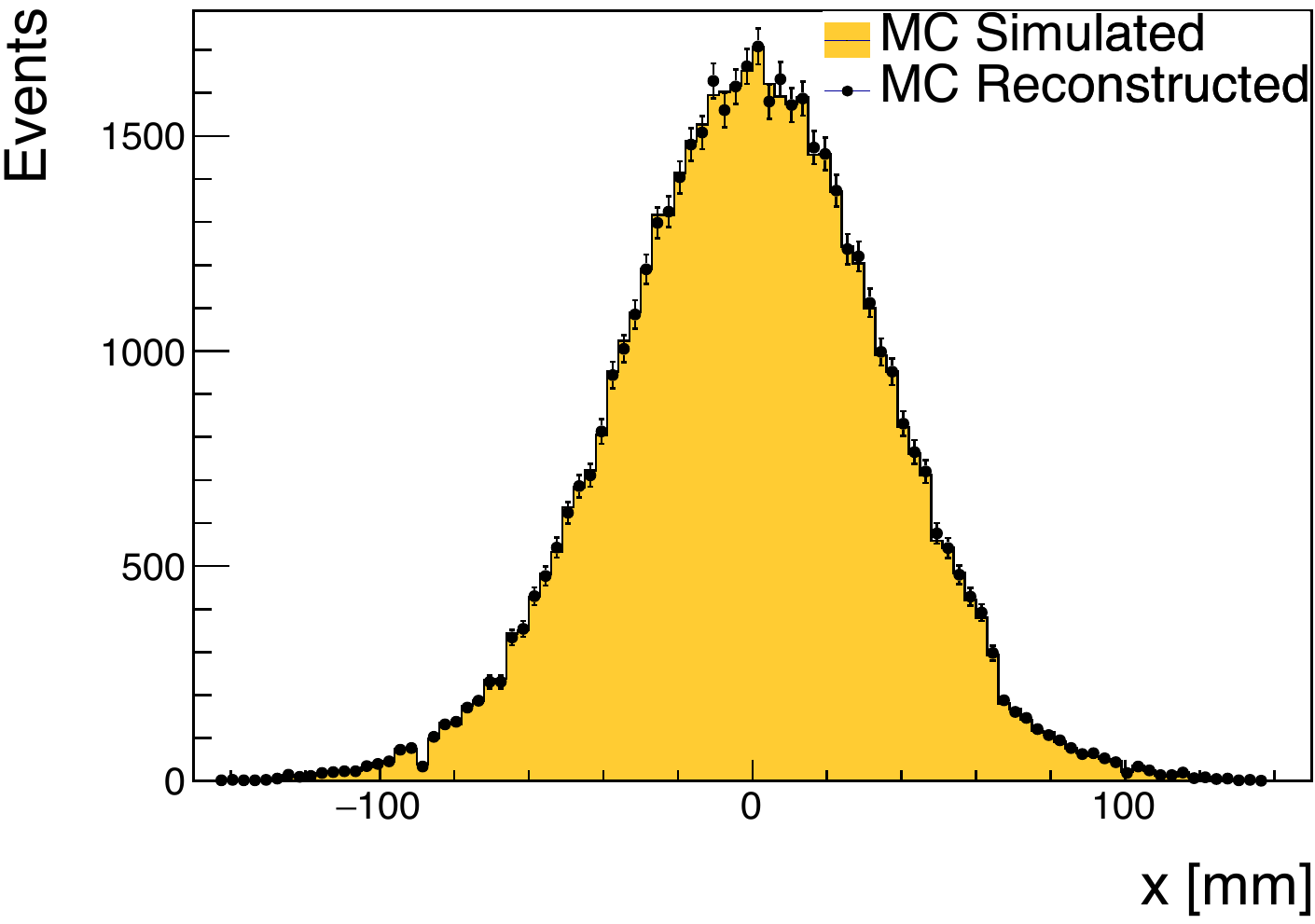}
\caption{}
\end{subfigure}
\begin{subfigure}[t]{0.495\textwidth}
\centering
\includegraphics[width=0.99\textwidth]{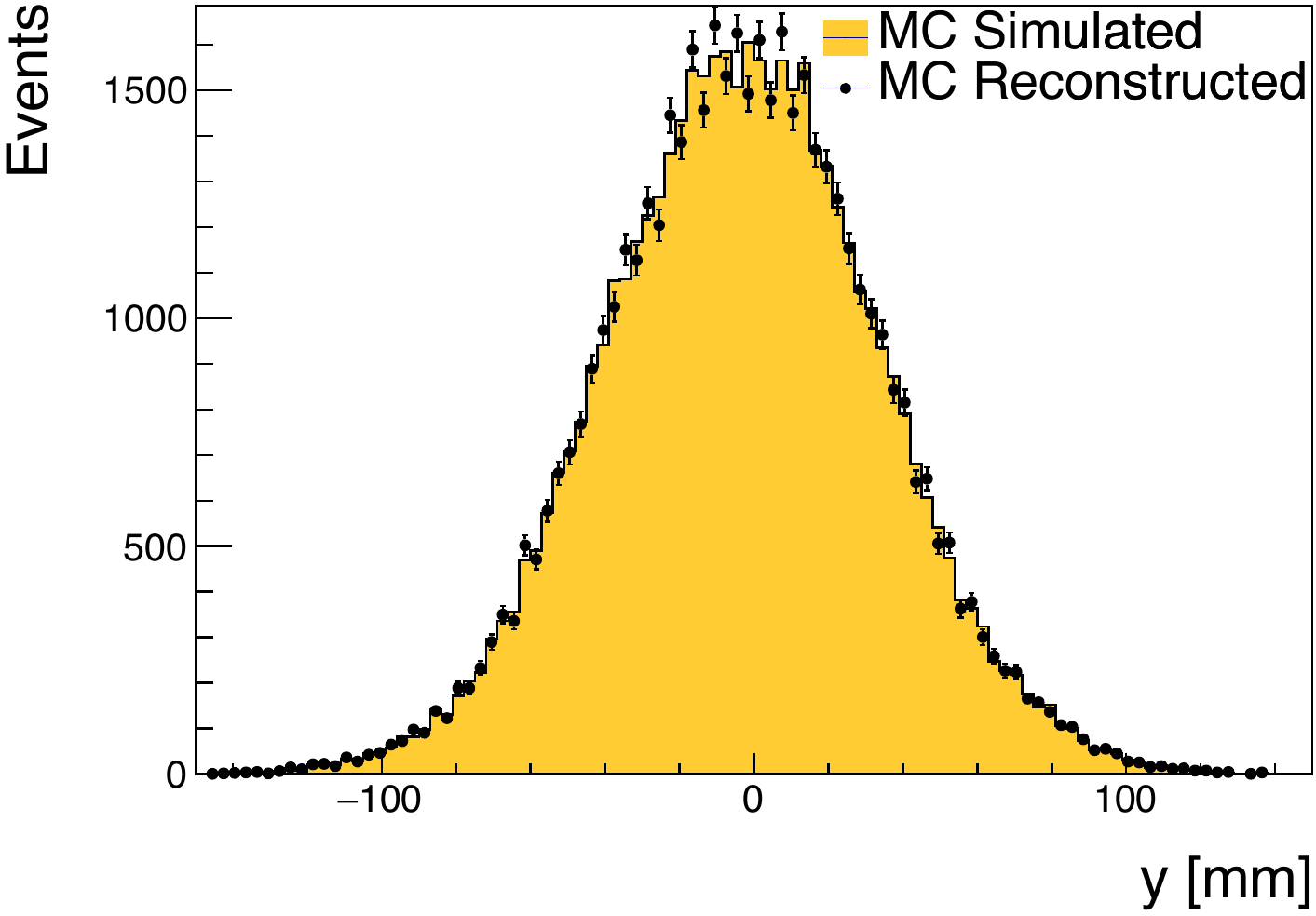}
\caption{}
\end{subfigure}

\begin{subfigure}[t]{0.495\textwidth}
\centering
\includegraphics[width=0.99\textwidth]{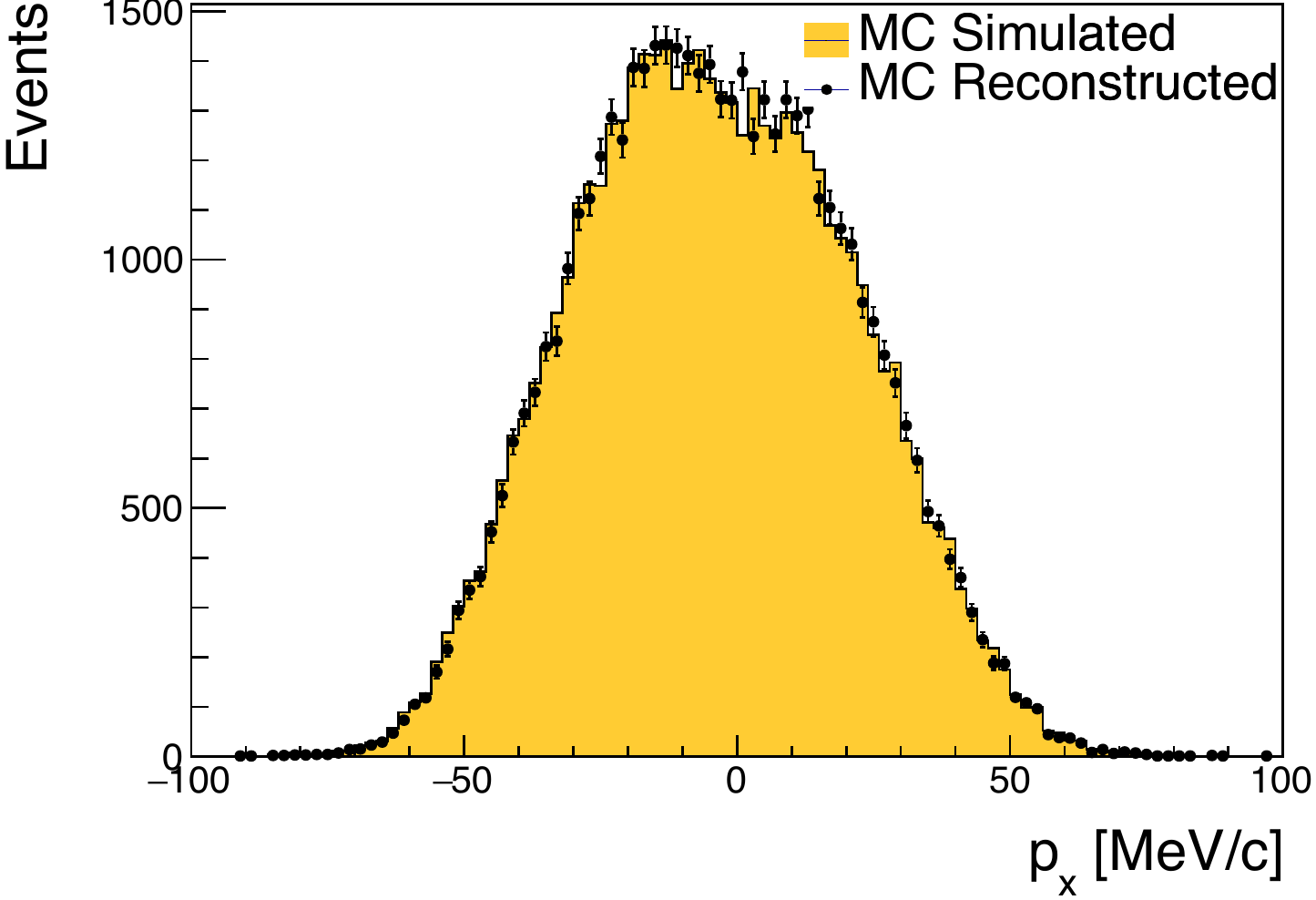}
\caption{}
\end{subfigure}
\begin{subfigure}[t]{0.495\textwidth}
\centering
\includegraphics[width=0.99\textwidth]{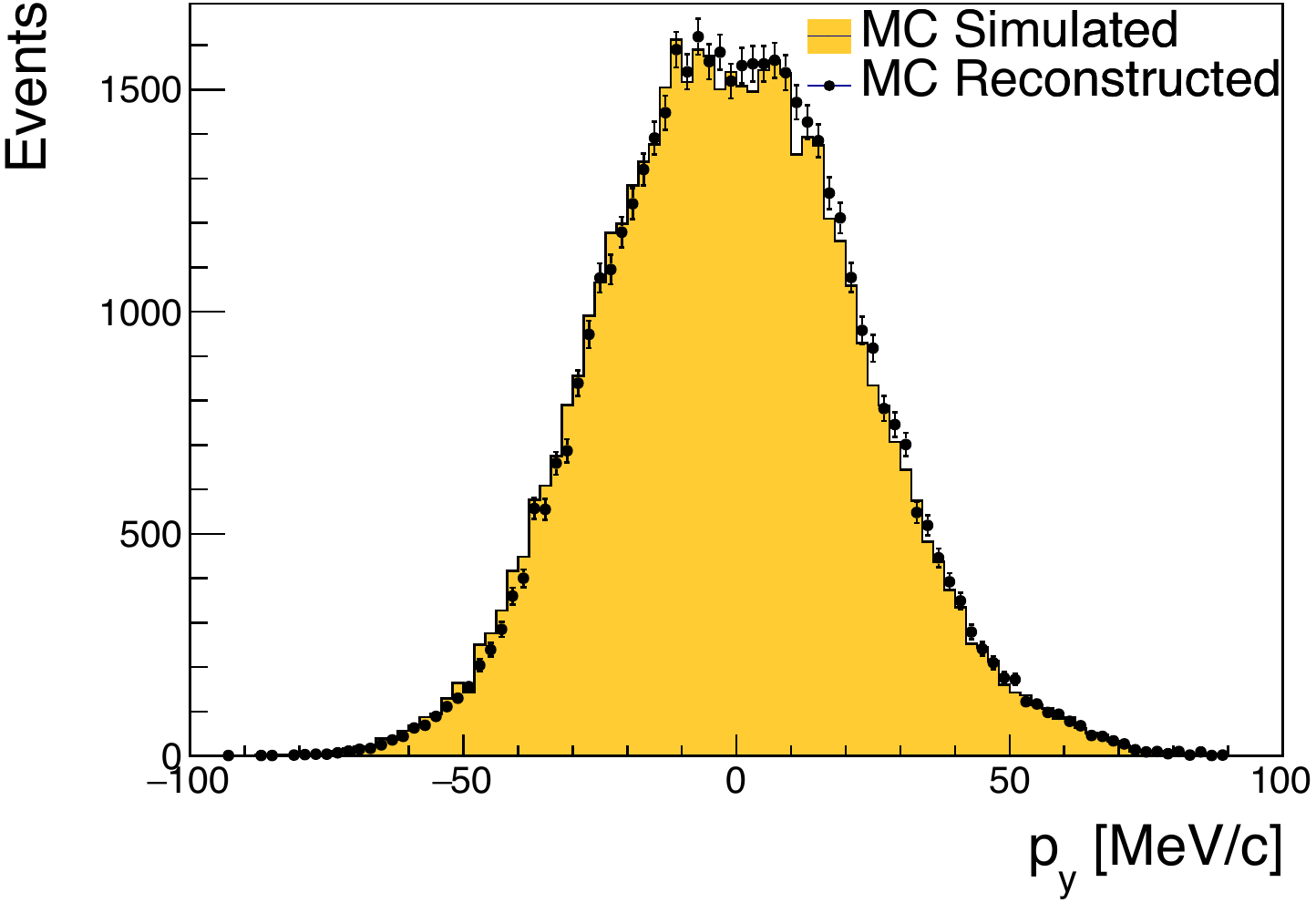}
\caption{}
\end{subfigure}

\caption{Position and momentum distributions of muons reconstructed at upstream
tracker station nearest to the absorber: a) $x$, b) $y$, c) $p_x$, d) $p_y$. The yellow histograms represent true MC simulations, and the markers represent the MC sample reconstructed using MAUS.}
\label{fig:tracker-posmom}
\end{figure*}

\subsection{KL calorimeter}
Hit-level reconstruction of the KL is  implemented in MAUS.  Individual PMT hits are unpacked from the DAQ or simulated from MC and the reconstruction associates them to identify the slabs that were hit and calculates the charge and charge-product corresponding to each slab hit. The KL has been used successfully to estimate the pion contamination in the MICE muon beamline~\cite{PionContaminationJINST}. 

\subsection{Electron-muon ranger}
Hit-level reconstruction of the EMR is  implemented in MAUS. The integrated ADC count and time over threshold are calculated for each bar that was hit.
The EMR reconstructs a wide range of variables that can be used for particle identification and momentum reconstruction. The  software and  performance of the EMR are described in detail in \cite{EMRJINST}. 

\subsection{Cherenkov}
The CKOV reconstruction takes the raw flash-ADC data, subtracts pedestals, calculates the charge and applies calibrations to determine the photo-electron yield. 

\subsection{Global reconstruction}
The aim of the Global Reconstruction is to take the reconstructed outputs from individual detectors and tie them together to form a global track. A likelihood for each particle hypothesis is also calculated. 

\subsubsection{Global track matching}

Global track matching is performed by collating particle hits (TOFs 0, 1 and 2, KL, Ckovs) and tracks (Trackers and EMR) from each detector using their individual reconstruction and combining them using a RK4 method to propagate particles between these detectors.The tracking is performed outwards from the cooling channel -- $\emph{i.e.}$, from the upstream tracker to the TOF0 detector, and from the downstream tracker to the EMR detector.
Track points are matched to form tracks using an RK4 method. Initially this is done independently for the upstream and downstream sections (i.e., either side of the absorber). As the trackers provide the most accurate position reconstruction, they are used as starting points for track matching, propagating hits outwards into the other detectors and then comparing the propagated position to the measured hit in the detector. The acceptance criterion for a hit belonging to a track is an agreement within the detector's resolution with an additional allowance for multiple scattering.  Track matching is currently performed for all TOFs, KL and EMR. 

The RK4 propagation requires the mass and charge of the particle to be known. Hence, it is necessary to perform track matching using a hypothesis for each particle type (muons, pions, and electrons). Tracks for all possible PID hypotheses are then passed to the Global PID algorithms. 

\subsubsection{Global PID} 

Global particle identification in MICE typically requires the combination of several detectors. The time-of-flight between TOF detectors can be used to calculate velocity, which is compared 
with the momentum measured in the trackers to identify the particle type. For all events but those with very low transverse momentum ($p_t$), charge can be determined from the direction of helical motion in the trackers. Additional information can be obtained from the CKOV, KL and EMR detectors. The global particle identification framework is designed to tie this disparate information into a set of hypotheses of particle types, with an estimate of the likelihood of each hypothesis.

The Global PID in MAUS uses a log-likelihood method to identify the particle species of a global track. It is based upon a framework of PID variables. Simulated tracks are used to produce probability density functions (PDFs) of the PID variables. These are then compared with the PID variables for tracks in real data to obtain a set of likelihoods for the PIDs of the track.

The input to the Global PID is several potential tracks from global track matching. During the track matching stage, each of these tracks was matched for a specific particle hypothesis. The Global PID then takes each track and determines the most likely PID following a series of steps:

\begin{enumerate}
\item Each track is copied into an intermediate track;
\item For each potential PID hypothesis $p$, the log-likelihood is calculated using the PID variables;
\item The track is assigned an object containing the log-likelihood for each hypothesis; and
\item From the log-likelhoods, the confidence level, C.L., for a track having a PID $p$ is calculated and the PID is set to the hypothesis with the best C.L.
\end{enumerate}

\subsection{Online reconstruction} 

During data taking, it is essential to visualize a detector's performance and have diagnostic tools to identify and debug unexpected behavior. This is accomplished through summary histograms of  high and low-level quantities from detectors. The implementation is through a custom multi-threaded application based on a producer--consumer pattern with thread-safe FIFO buffers. Raw data produced by the DAQ are streamed through a network and consumed by individual detector \textit{mappers} described in Section 3. The reconstructed outputs produced by the \textit{mappers},  are in turn consumed by the \textit{reducers}. The \textit{mappers} and \textit{reducers} are distributed among the threads to balance the load. Finally,  outputs from the \textit{reducers} are written as histogram images. Though the framework for the online reconstruction is based on parallelized processing of spills, the reconstruction modules are the same as those used for offline processing. A lightweight tool based on Django \cite{Django} provides live web-based visualization of the histogram images as and when they are created. Typical data rates during experimental operations were $\sim$ 300 MB/s. The average event rate varied, depending on the configuration of the beamline, with the maximum instantaneous rate being $\sim$ 150 kHz. MAUS performance matched the data rates and  online reconstruction happened virtually ``live'' with the reconstructed outputs  available instantly allowing collaborators to monitor the quality of the data being acquired.

 \section{Summary}

The MICE collaboration has developed the MAUS software suite to simulate the muon beamline, simulate the MICE detectors, and reconstruct both simulated and real data. The software also provides global track-matching and particle-identification capabilities. Simplified programming interfaces and testing environments enable productive development. MAUS has been successfully used to reconstruct data online during data collection. In addition, MAUS is routinely used to perform reconstruction of the entire MICE data volume on batch production systems. MICE has collected $\sim$ 15 TB of raw data and a full reconstruction of the data is performed with each released version of MAUS. The batch systems are also used to perform compute-intensive simulations with various configurations of the beamline and the cooling channel. 

\acknowledgments

The work described here was made possible by grants from Department of Energy and National Science Foundation
(USA), the Instituto Nazionale di Fisica Nucleare (Italy), the Science and Technology Facilities Council
(UK), the European Community under the European Commission Framework Programme 7 (AIDA project,
grant agreement no. 262025, TIARA project, grant agreement no. 261905, and EuCARD), the Japan Society
for the Promotion of Science and the Swiss National Science Foundation, in the framework of the SCOPES
programme. We gratefully acknowledge all sources of support. We are grateful to the support given to us
by the staff of the STFC Rutherford Appleton and Daresbury Laboratories. We acknowledge the use of Grid
computing resources deployed and operated by GridPP \cite{GridPP} in the UK.

\bibliography{mice}                                                                           

\begin{thebibliography}{10}

\bibitem{CoolingIPAC18}
{MICE Collaboration: T. Mohayai et al.}
\newblock {First Demonstration of Ionization Cooling in MICE}.
\newblock In {\em Proc. 2018 International Particle Accelerator Conference,
  Vancouver}, 2018.
\newblock FRXGBE3,
  \href{https://doi.org/10.18429/JACoW-IPAC2018-FRXGBE3}{https://doi.org/10.18429/JACoW-IPAC2018-FRXGBE3}.

\bibitem{IDR}
{The IDS-NF collaboration}.
\newblock International design study for the neutrino factory: Interim design
  report, 2011.
\newblock IDS-NF-020,
  \href{https://www.ids-nf.org/wiki/FrontPage/Documentation}{https://www.ids-nf.org/wiki/FrontPage/Documentation}.

\bibitem{MC_Overview}
Steve Geer.
\newblock Muon {C}olliders and {N}eutrino {F}actories.
\newblock {\em Annual Review of Nuclear and Particle Science}, 59:345 -- 367,
  2009.

\bibitem{BeamlineJINST}
M.~Bogomilov et~al.
\newblock {The MICE Muon Beam on ISIS and the beam-line instrumentation of the
  Muon Ionization Cooling Experiment}.
\newblock {\em JINST}, 7:P05009, 2012.

\bibitem{BeamCharacterisationEurPhysJ}
D.~Adams et~al.
\newblock {Characterisation of the muon beams for the Muon Ionisation Cooling
  Experiment}.
\newblock {\em Eur. Phys. J.}, C73(10):2582, 2013.

\bibitem{NIMA_TOF}
R.~Bertoni et~al.
\newblock The design and commissioning of the {MICE} upstream time-of-flight
  system.
\newblock {\em {Nucl. Instrum. Meth.}}, A615:14 -- 26, 2010.

\bibitem{KLOE}
{KLOE} Collaboration.
\newblock The {KLOE} electromagnetic calorimeter.
\newblock {\em Nucl. Instrum. Meth.}, A494:326 -- 331, 2002.

\bibitem{KLOE2}
F.~Ambrosino et~al.
\newblock Calibration and performances of the {KLOE} calorimeter.
\newblock {\em Nucl. Instrum. Meth.}, A598:239 -- 243, 2009.

\bibitem{PionContaminationJINST}
D.~Adams et~al.
\newblock {Pion Contamination in the MICE Muon Beam}.
\newblock {\em JINST}, 11(03):P03001, 2016.

\bibitem{EMRJINST}
D.~Adams et~al.
\newblock {Electron-muon ranger: performance in the MICE muon beam}.
\newblock {\em JINST}, 10(12):P12012, 2015.

\bibitem{EMRJINST11}
R.~Asfandiyarov et~al.
\newblock {The design and construction of the MICE Electron-Muon Ranger}.
\newblock {\em JINST}, 11(10):T10007, 2016.

\bibitem{FirstSingleParticle}
D.~Adams et~al.
\newblock {First particle-by-particle measurement of emittance in the Muon
  Ionization Cooling Experiment}.
\newblock {\em Eur. Phys. J.}, C79(3):257, 2019.

\bibitem{CkovIEEE}
{L. Cremaldi, D. A. Sanders, P. Sonnek, D. J. Summers, and J. Reidy, Jr}.
\newblock A cherenkov radiation detector with high density aerogels.
\newblock {\em {IEEE Trans. Nucl. Sci.}}, 56:1475--1478, 2009.

\bibitem{TrackersNIM}
M.~Ellis, P.R. Hobson, P.~Kyberd, J.J. Nebrensky, A.~Bross, et~al.
\newblock {The Design, construction and performance of the MICE scintillating
  fibre trackers}.
\newblock {\em Nucl. Instrum. Meth.}, A659:136--153, 2011.

\bibitem{SCONS}
\href{https://scons.org}{https://scons.org}.

\bibitem{bazaar}
\href{https://bazaar.canonical.com}{https://bazaar.canonical.com}.

\bibitem{launchpad}
\href{https://launchpad.net}{https://launchpad.net}.

\bibitem{ROOT}
R.~Brun and F.~Rademakers.
\newblock Root - an object oriented data analysis framework.
\newblock {\em Nucl. Instrum. Meth.}, A389:81 -- 86, 1997.
\newblock \href{https://root.cern.ch}{https://root.cern.ch}.

\bibitem{GEANT4}
S~Agnostinelli et~al.
\newblock Geant4 {-} a simulation toolkit.
\newblock {\em Nucl. Instrum. Meth.}, A506:250 -- 303, 2003.

\bibitem{scilinux}
\href{https://scientifixlinux.org}{https://scientificlinux.org}.

\bibitem{centos}
\href{https://centos.org}{https://centos.org}.

\bibitem{fedora}
\href{https://getfedora.org}{https://getfedora.org}.

\bibitem{ubuntu}
\href{https://ubuntu.com}{https://ubuntu.com}.

\bibitem{MapReduce}
J.~Dean and S.~Ghemawat.
\newblock {MapReduce}: Simplified data processing on large clusters.
\newblock In {\em Proceedings of {OSDI}04}, 2004.
\newblock
  \href{http://research.google.com/archive/mapreduce.html}{http://research.google.com/archive/mapreduce.html}.

\bibitem{JSON}
\href{https://json.org}{https://json.org}.

\bibitem{JSONCPP}
\href{https://github.com/open-source-parsers/jsoncpp}{https://github.com/open-source-parsers/jsoncpp}.

\bibitem{Jenkins}
\href{https://jenkins-ci.org}{https://jenkins-ci.org}.

\bibitem{G4Beamline}
Thomas~J. Roberts and Daniel~M. Kaplan.
\newblock {G4BeamLine} programme for matter dominated beamlines.
\newblock In {\em Proc. 2007 Particle Accelerator Conference, Albuquerque},
  2007.
\newblock THPAN103.

\bibitem{ICOOL}
R.~C. Fernow.
\newblock Icool: A simulation code for ionization cooling of muon beams.
\newblock In {\em Proc. 1999 Particle Accelerator Conference, New York}, 1999.

\bibitem{GEANT4HPhysics}
J~Apostolakis et~al.
\newblock {Geometry and physics of the geant4 toolkit for high and medium
  energy applications}.
\newblock {\em Radiation Physics and Chemistry}, 78(10):859 -- 873, 2009.

\bibitem{GDML}
J.~McCormick R.~Chytracek.
\newblock Geometry description markup language for physics simulation and
  analysis applications.
\newblock {\em IEEE Trans. Nucl. Sci.}, 53:2892--2896, 2006.

\bibitem{fastrad}
\href{https://fastrad.net}{https://fastrad.net}.

\bibitem{xslt}
\href{https://www.w3.org/standards/xml/transformation}{https://www.w3.org/standards/xml/transformation}.

\bibitem{PennEnvelope}
G.~Penn.
\newblock Beam envelope equations in a solenoidal field.
\newblock {\em Phys. Rev. Lett.}, 85:764 -- 767, 2000.

\bibitem{NonlinearNote}
R.D. Ryne and C.~Rogers.
\newblock Nonlinear effects in the {MICE} step iv lattice.
\newblock {MICE} {N}ote 461,
  \href{http://mice.iit.edu/micenotes/public/pdf/MICE0461/MICE0461.pdf}{http://mice.iit.edu/micenotes/public/pdf/MICE0461/MICE0461.pdf},
  2015.

\bibitem{MiddletonThesis}
Sophie~Charlotte Middleton.
\newblock {\em {Characterisation of the MICE experiment}}.
\newblock PhD thesis, Imperial College London, 2018.

\bibitem{MaryLie}
{R.D. Ryne et al.}
\newblock {Recent progress on the MARYLIE/IMPACT beam dynamics code}.
\newblock In {\em Proc. 9th International Computational Accelerator Physics
  Conference (ICAP 06)}, 2006.
\newblock
  \href{http://icap2006.web.cern.ch/icap2006/}{http://icap2006.web.cern.ch/icap2006/}.

\bibitem{MAUSNuFact}
C.T Rogers et~al.
\newblock Muon front end for the neutrino factory.
\newblock {\em Phys. Rev. ST Accel. Beams}, 16:040104, 2013.

\bibitem{TrackerSoftwareJINST}
A.~Dobbs, C.~Hunt, K.~Long, E.~Santos, M.A. Uchida, P.~Kyberd, C.~Heidt,
  S.~Blot, and E.~Overton.
\newblock The reconstruction software for the mice scintillating fibre
  trackers.
\newblock {\em JINST}, 11(12):T12001, 2016.

\bibitem{Django}
\href{https://www.djangoproject.com}{https://www.djangoproject.com}.

\bibitem{GridPP}
\href{https://www.gridpp.ac.uk}{https://www.gridpp.ac.uk}.

\end{thebibliography}
\bibliographystyle{unsrt}

\end{document}